\documentclass[10pt,twocolumn]{article}

\usepackage{amsmath}
\usepackage{amssymb}
\usepackage[margin=0.8in]{geometry}
\usepackage{xcolor}
\usepackage{graphicx}
\usepackage{hyphenat}
\usepackage{bm}
\usepackage{siunitx}
\usepackage{enumerate}
\usepackage{abstract}

\usepackage[superscript]{cite}

\usepackage{draftwatermark}
\SetWatermarkText{\bf PREPRINT ~~~ PREPRINT ~~~ PREPRINT   ~~~ PREPRINT}
\SetWatermarkFontSize{6cm}
\SetWatermarkAngle{90}
\SetWatermarkHorCenter{20.5cm}
\SetWatermarkColor[gray]{0.75}

\date{\textbf{Accepted for publication, 2022}}

\title{14-moment maximum-entropy modelling of collisionless ions for Hall thruster discharges}

\author{Stefano Boccelli\thanks{Corresponding author: stefano.boccelli@polimi.it}\\
\textit{\normalsize University of Ottawa, ON, Canada,}\\
\textit{\normalsize and Politecnico di Milano, Milan, Italy,}\\
\textit{\normalsize and von Karman Institute for Fluid Dynamics, Sint-Genesius-Rode, Belgium.}\\[2ex]
James G.~McDonald\\
\textit{\normalsize University of Ottawa, ON, Canada.}\\[2ex]
Thierry E.~Magin\\
\textit{\normalsize von Karman Institute for Fluid Dynamics, Sint-Genesius-Rode, Belgium.}
}

\begin{document}

\twocolumn[
  \begin{@twocolumnfalse}

    \maketitle

    \begin{abstract}
    \noindent Ions in Hall effect thrusters are often characterized by a low collisionality.
    In the presence of acceleration fields and azimuthal electric field waves, this results in strong deviations
    from thermodynamic equilibrium, introducing kinetic effects.
    This work investigates the application of the 14-moment maximum-entropy model to this problem.
    This method consists in a set of 14 PDEs for the density, momentum, pressure tensor components, heat flux vector and fourth-order moment
    associated to the particle velocity distribution function.
    The model is applied to the study of collisionless ion dynamics in a Hall thruster-like configuration,
    and its accuracy is assessed against different models, including the Vlasov kinetic equation.
    Three test cases are considered: a purely axial acceleration problem, the problem of ion-wave trapping and finally the evolution of ions in the axial-azimuthal plane.

    Most of this work considers ions only, and the coupling with electrons is removed by prescribing
    reasonable values of the electric field.
    This allows us to obtain a direct comparison among different ion models.
    However, the possibility to run self-consistent plasma simulations is also briefly discussed,
    considering quasi-neutral or multi-fluid models.
    The maximum-entropy system appears to be a robust and accurate option for the considered test cases.
    The accuracy is improved over the simpler pressureless gas model (cold ions) and the Euler equations for gas dynamics, while the computational
    cost shows to remain much lower than direct kinetic simulations.
    \end{abstract}

    \vskip2ex
  \end{@twocolumnfalse}
]

\saythanks


\section{Introduction}

Hall effect thrusters \cite{morozov2000fundamentals,zhurin1999physics} are  increasingly popular space propulsion devices.
Presently, they are widely employed for satellite maneuvering and are being envisaged for future manned space transportation.
Despite the sub-Newton levels of thrust offered by these devices, their high specific impulse makes them an appealing alternative 
to traditional chemical propulsion,\cite{goebel2008fundamentals} and makes them almost mandatory for long-distance missions.\cite{racca2001capability}

Hall thrusters are $\bm{E}\times\bm{B}$ plasma devices,\cite{boeuf2017tutorial}
where an externally imposed magnetic field, $\bm{B}$, is responsible for reducing 
the electron mobility and thus creating an axial electric field, $\bm{E}$, that ultimately accelerates ions.
The maximum value of the magnetic field is often in the order of $|\bm{B}_{\mathrm{max}}| \approx 200~\si{G}$, 
that is designed such that the ion trajectory is approximately unaltered.
In this work, we consider ions to be completely unmagnetized.
In real Hall thrusters, the electric field is not strictly axial, but is affected by the actual three-dimensional thruster geometry, and 
often shows travelling waves originating from plasma instabilities.


Ions in Hall thruster discharges are often characterized by a low collisionality, both among themselves and with other species (electrons and background neutrals).
Inside the thruster channel, the electric field is often strong enough for the residual effect of ion collisions to be small:
under the effects of the acceleration field and of the ionization source terms, the ion velocity distribution function (VDF) 
inside the thruster often assumes non-equilibrium shapes.\cite{boccelli2020collisionless,chaplin2018laser, mazouffre2010spatio}
On the other hand, as ions leave the thruster channel and enter the plume, 
charge exchange (CEX) and momentum exchange (MEX) ion-neutral collisions play an increasingly important role.\cite{mikellides2005elastic}

Modelling ions with a full accuracy can be done by solving the kinetic equation,\cite{ferziger1972mathematical} with either deterministic or particle-based numerical 
approaches.\cite{birdsall2004plasma}
The high dimensionality of this equation makes kinetic solvers usually computationally expensive. 
Nonetheless, kinetic solvers are often employed when simulating ions in Hall thruster plasmas, since their computational cost remains much smaller than
the expense associated to the simulation of electrons.\cite{parra2006two,hara2012one,shashkov2017one}

As opposed to kinetic methods, one can opt for a fluid-like model. 
Such models are drastic simplifications of the kinetic problem and consist in solving only for a limited set of moments of the VDF.
The simplest fluid model probably consists in the pressureless gas equations.\cite{chen2003formation}
This model has been frequently employed in the plasma propulsion literature. \cite{ahedo2002model,yim2008computational,choi2008modeling,de20093d}
Indeed, this model has proven to be very effective in reproducing the lower order moments of the VDF, such as the density and velocity.
However, such a model does not give any information on the ion temperature, that may be an important quantity in a number of scenarios.\cite{lafleur2016theory2}
The Euler equations of gas dynamics significantly improve things by adding an energy equation.
Still, these equations remain an approximation and one should consider that, due to the low collisionality and the electrical acceleration, 
the solution is likely to show temperature anisotropy.
Moreover, the distribution function can show strong asymmetries associated to a non-zero heat flux 
(thus breaking the adiabatic assumption of the Euler equations).
Higher-order moments can also show non-equilibrium effects. 
The Navier-Stokes equations extend the validity of the Euler equations and include a heat flux, but are still based on small perturbations of 
a local Maxwellian VDF and therefore are not considered here, since we wish instead to obtain a formulation that allows for 
strong departures from equilibrium.

A variety of moment systems have been proposed over the years to describe non-equilibrium gases, with varying degrees of success.
Among a number of possibilities, we should mention the Grad method \cite{grad1949kinetic} and the Quadrature-based moment methods.\cite{fox2009higher}
Such methods allow one to build approximations of the kinetic equation up to an arbitrary order and are expected to asymptotically recover the kinetic solution, as the number of considered moments is increased.
The number of moments that is required to obtain a given accuracy is not easily estimated \cite{torrilhon2015convergence} and depends on the considered test case.
Besides the applications to neutral gases, moment methods have often been employed to the study of plasmas. \cite{miller2016multi,lilly2008regions,dong2019global}
In this work, we consider the maximum-entropy class of moment methods.\cite{levermore1996moment,muller2013rational}
Such models have proven to be robust even in strongly non-equilibrium conditions, showing a non-negative VDF by construction.
Moreover, the maximum-entropy formulation guarantees that the resulting system of governing equations is hyperbolic (with clear numerical advantages over parabolic systems),
while at the same time providing in a natural way with pressure anisotropy and a (non-local) heat flux.
We consider here a fourth-order maximum-entropy method, that results in 14 governing equations. 
For this 14-moment system, the formulation of approximated interpolative solutions to the entropy-maximisation problem has solved the long lasting issue 
of a high computational cost, and made a class of such methods computationally affordable.\cite{mcdonald2013affordable,mcdonald2016approximate,giroux2021approximation}
To date, maximum-entropy methods have been applied to a number of single and multi-fluid problems,\cite{tensuda2015application,forgues2019higher}
microfluidics,\cite{mcdonald2011extended} radiation transport\cite{sarr2020second} and were recently applied to the study of electrons in the magnetosphere\cite{ng2018using} and in $\bm{E}\times\bm{B}$ low-temperature plasma discharges.\cite{boccelli202014}


In this work, we investigate the application of the 14-moment fourth-order maximum-entropy method to the description of 
collisionless unmagnetized ions in Hall thruster-like discharges.
We focus on the thruster channel and neglect the plume.
In particular, we aim at studying the process of axial acceleration of ions in the presence of azimuthally travelling waves.
For this problem, we wish to assess how accurately the maximum-entropy method can recover the kinetic solution, and to compare
it to simpler fluid models.
For this reason, only ions are considered in most of this work, and the coupling with electrons (often modeled through the Poisson equation) 
is completely disregarded.
Instead, we prescribe here some reasonable profiles for the electric field and for the ionization rate.
In this way, we artificially remove the need to simulate electrons and thus remove all associated modelling
uncertainties.
Our results do not represent full plasma simulations, but are instead comparisons of the different modelling strategies.
However, some guidelines for embedding this model into full plasma simulations, as well as some practical examples, are also given.

The maximum-entropy method is introduced in Section~\ref{sec:max-ent}, and the governing equations are discussed,
together with the source terms and the closure for higher order moments.
Then, the method is applied to Hall thruster problems.
First, in Section~\ref{sec:axial-accel}, we consider a one-dimensional domain located at the channel centerline and directed axially.
This allows us to assess the accuracy of the fourth-order maximum-entropy method for the problem of production and steady 
axial acceleration of ions.
Besides this problem, Hall thrusters often show azimuthally travelling waves:
in Section~\ref{sec:ion-wave-trapping} we consider a one-dimensional periodic domain, oriented along the azimuthal direction, and 
located at the thruster exit.
A travelling electric field wave is imposed, with a reasonable frequency and phase velocity, and is observed to cause ion-wave trapping.
This test case allows us to discuss in detail the computational cost of the maximum-entropy method, that is compared to the cost of the classical 
fluid and kinetic solutions in Section~\ref{sec:computational-efficiency}.
In Section~\ref{sec:full-test-case}, the two previous test cases are combined, and we consider the ion evolution on a 
two-dimensional axial-azimuthal plane.
Finally, Section~\ref{sec:self-consistent} discusses how the maximum-entropy system can be supplemented by an electron model, 
to run either quasi-neutral or multi-fluid plasma simulations.


\section{The 14-moment maximum-entropy model}\label{sec:max-ent}

The moment equations can be formally obtained starting from the kinetic equation,\cite{ferziger1972mathematical}
discussed in the following.


\subsection{The kinetic model}

For collisionless and unmagnetized ions, we write
\begin{equation}\label{eq:kinetic-eq}
  \frac{\partial f}{\partial t} + \bm{v} \cdot \frac{\partial f}{\partial \bm{x}} + \frac{q \bm{E}}{m} \cdot \frac{\partial f}{\partial \bm{v}} = C_{\mathrm{iz}}(\bm{x},\bm{v}) \, ,
\end{equation}

\noindent where $f=f(\bm{x},\bm{v},t)$ is the VDF at position $\bm{x}$ and velocity $\bm{v}$, while $q$ and $m$ are the ion charge and mass.
The source term $C_{\mathrm{iz}}(\bm{x},\bm{v})$ models the chemical production of ions.
The only ionization mechanism considered here is direct ionization of neutral particles by electron impact.
Given the large mass ratio between electrons and neutrals, one can neglect the transfer of momentum and 
assume that ions are produced with a velocity distribution that follows the velocity distribution of the background 
neutrals:
\begin{equation}
  C_{\mathrm{iz}}(\bm{x},\bm{v}) = C_{\mathrm{iz}}(\bm{x}) \, g(\bm{v}; T_n, \bm{u}_n) \, ,
\end{equation}

\noindent where $g$ is a normalized Maxwellian distribution at the neutral temperature, $T_n$, and average velocity, $u_n$.
The parameter $C_{\mathrm{iz}}(\bm{x})$ represents the local ionization rate, measured in $[\mathrm{ions}/\si{m^3 s}]$.
In principle, this term should depend on the neutral density, and on the local electron density and temperature, 
and may be affected by electron non-equilibrium.\cite{fedotov1999electron,shagayda2017analytic}
However, since in this work we are not interested in solving a fully coupled problem,
the ion source term is here prescribed.

The moments of the VDF, $f$, are obtained as integrals of $f$ over the velocity space.
We employ the notation
\begin{equation}
  M_{\psi} = \left< \psi \right> = \iiint_{-\infty}^{+\infty} \psi f \, \mathrm{d}^3 v \, ,
\end{equation}

\noindent where $\psi$ is a particle quantity (such as its mass $m$ or momentum $m\bm{v}$), and $M_{\psi}$ is the 
corresponding moment.
The density, momentum and pressure tensor components read 
\begin{equation}
  \rho = \left< m \right> \ \ , \ \ \ \rho u_i = \left< m v_i \right> \ \ , \ \ \ P_{ij} = \left< m c_i c_j\right> \, ,
\end{equation}

\noindent with $c_i = v_i - u_i$ the peculiar velocity in the direction $i=\{x,y,z\}$.
The heat-flux tensor components and the fourth-order moment read 
\begin{equation}
  Q_{ijk} = \left< m c_i c_j c_k \right> \ \ , \ \ \ R_{iijj} = \left< m c_i c_i c_j c_j \right> \, ,
\end{equation}

\noindent where repeated indices indicate summation, and the heat-flux vector is indicated here as $q_i = Q_{ikk}$.
Notice that this definition differs from the common fluid dynamic definition by a factor of $1/2$.
The fourth-order moment $R_{iijj}$ is obtained as the average of the VDF weighted by the peculiar velocity to
the fourth-power.
This moment gives an indication of the thickness of the tails of the VDF.
Flat-top distributions such as the Druyvestein's show lower values for $R_{iijj}$ with respect to a Maxwellian,
since less particles are located in the tails and more particles are in the bulk instead.
From a direct integration, the equilibrium value for $R_{iijj}$ is seen to be $R_{iijj}^{\mathrm{eq}}=15\,P^2/\rho$, 
where $P=(P_{xx}+P_{yy}+P_{zz})/3$ is the hydrostatic pressure.

One convenient way to assess non-equilibrium is to analyze the value of dimensionless moments (superscript $\star$),
that are obtained by dividing the dimensional moments by the density and by suitable powers of the characteristic thermal velocity $(P/\rho)^{1/2}$:
\begin{equation}\label{eq:dimensionless-moments-def}
  \begin{cases}
  \begin{aligned}
  \rho^\star    &= 1 , \\ 
  P_{ij}^\star  &= P_{ij}/P , \\
  Q_{ijk}^\star &= Q_{ijk}\Big/\left[ \rho \left(P/\rho\right)^{3/2} \right], \\ 
  R_{iijj}^\star &= R_{iijj}\Big/\left[ \rho \left(P/\rho\right)^{2} \right]. \\ 
  \end{aligned}
  \end{cases}
\end{equation}

\noindent These dimensionless moments will be used in the next sections for plotting the solution in (dimensionless) moment space.
At equilibrium, $P^\star_{ij} = \delta_{ij}$, $Q^\star_{ijk} =0$ and $R_{iijj}^\star=15$.


\subsection{The maximum-entropy assumption}

By taking moments of the kinetic equation, one obtains an infinite hierarchy of moment equations.\cite{ferziger1972mathematical}
This hierarchy can be truncated at the desired order, to obtain a finite and manageable set of PDEs.
A a result of the truncation, a number of unknown moments are left in the fluxes, and require to introduce some closure assumption.
For instance, the Euler equations are obtained by considering equations up to the second order in the particle velocity (energy equation)
and by assuming that all closing moments (heat flux and pressure deviator) are zero.
This choice corresponds to assuming that the VDF is Maxwellian.
In the present case, we will consider $N=14$ PDEs (see below) and write equations for moments up to the 
fourth-order in the particle velocity.

Whatever the chosen truncation, if one knew the expression of the VDF associated to the $N$ known moments, 
the closing moments would be easily computed by direct integration.
However, finding such a VDF is not a well defined task, since there may be an infinite number of VDFs that are compatible
with a finite number of moments.
The maximum-entropy method defines a strategy for selecting a VDF:\cite{levermore1996moment}
among every possible $f$ compatible with the given $N$ moments, one selects the one that maximizes the entropy 
\begin{equation}\label{eq:entropy-formula}
  \mathsf{S} = -k_B \iiint f \, \ln \frac{f}{y} \, \mathrm{d}^3 v \, ,
\end{equation}

\noindent written here for a classical non-degenerate gas, where $y$ is a normalizing constant 
and $k_B$ the Boltzmann constant.
It is important to stress that, in the maximum-entropy moment method formulations, we maximize the entropy under the constraint
that its moments have a specific value.
For this reason, the outcome of the procedure is not necessarily a Maxwellian distribution.
The Maxwellian is thus only a special result, that is obtained at equilibrium, 
while this formulation can span both equilibrium and non-equilibrium states.

The choice of selecting the maximum-entropy distribution, among all possible distributions, 
has a clear meaning in the presence of collisions.
Indeed, the Boltzmann collision operator is known to drive the VDF towards equilibrium and to vanish when the entropy, $\mathsf{S}$,
is at its maximum.\cite{ferziger1972mathematical}
However, completely regardless of collisionality, the maximum-entropy VDF has another important 
property: 
it can be shown to be the most likely distribution, among all possible distributions that integrate to the $N$ known moments.\cite{muller1985thermodynamics}
This gives the motivation for employing the maximum-entropy method in the present work, in total absence of collisions.

We consider here fourth-order maximum-entropy moment methods, where governing equations are written for moments up to
the fourth-order in the particle velocity.
Higher order moments only appear as closing terms.
The simplest fourth-order maximum-entropy system is composed of 14-moments and is discussed in the next section. 


\subsection{The 14-moment system of PDEs}

Maximum-entropy VDFs can be shown to be in the form\cite{levermore1996moment}
\begin{equation}
  f = \exp\left[\bm{\alpha}^\intercal \bm{m}(\bm{v}) \right] \, ,
\end{equation}

\noindent where $\bm{\alpha}$ is a vector of weights and the elements of $\bm{m}(\bm{v})$ are powers of the particle velocity, up
to the desired order.
In the present work, the VDF if fourth-order in the velocity and is described by 14 parameters:
%
\begin{equation}\label{eq:f14-definition}
  f_{14}=\exp(\alpha_0 + \alpha_i v_i + \alpha_{ij} v_i v_j + \alpha_{i,3}v_iv^2 + \alpha_4 v^4) \, .
\end{equation}

\noindent This VDF can represent strongly anisotropic and asymmetric VDFs, with non-Maxwellian kurthosis
\cite{boccelli202014,ng2018using}
and recovers, as a limiting case, the Maxwellian and the Druyvestein's distributions. 
Moreover, it is positive by construction, regardless of the degree of non-equilibrium.

The vector of coefficients, $\bm{\alpha}$, can be obtained explicitly by solving the entropy maximisation 
problem.\cite{dreyer1987maximisation}
Given the strong non-linearity introduced by the exponential function, 
there is no simple algebraic relation between the coefficients in $\bm{\alpha}$
and the fluid variables, except for specific states such as equilibrium.
To date, the most practical approach for solving the entropy maximization problem 
for a general set of moments and far from equilibrium consists in numerical optimization 
(most often, Newton iterations).
This approach quickly becomes prohibitively expensive if it is to be applied to multi-dimensional
fluid dynamic simulations, where one has to run the Newton iterations at every timestep and in every 
cell of the computational domain.
Hardware acceleration has often been employed to address this issue.\cite{schaerer2017efficient,garrett2015optimization}
As discussed in the following, in this work we employ instead a computationally affordable approximation to 
the entropy maximization problem, that allows us to solve this issue, 
by avoiding a direct computation of the coefficient vector, $\bm{\alpha}$.

For each velocity term appearing in $f_{14}$, one can compute a moment of the kinetic equation, Eq.~\eqref{eq:kinetic-eq}.
The final result is a system of 14 hyperbolic governing equations, that we write in balance-law form as

\begin{equation}\label{eq:governing-equations-fluid}
    \frac{\partial \bm{U}}{\partial t} + 
    \bm{\nabla} \cdot \bm{F} =
    \bm{S}_{\bm{E}} +
    \bm{S}_{\mathrm{iz}} \, ,
\end{equation}

\noindent with $\bm{U}$ the state vector, $\bm{F} = \bm{F}(\bm{U}) = [\bm{F}_x, \bm{F}_y, \bm{F}_z]$ the fluxes along $x$, $y$, and $z$.
The terms $\bm{S}_{\bm{E}}$ and $\bm{S}_{\mathrm{iz}}$ are the electrostatic and ionization source terms.
The state vector reads
\begin{equation}\label{eq:Ui-definitions}
  \bm{U} 
= 
  \begin{pmatrix}
    \left< m \right> \\
    \left< m v_x \right> \\
    \left< m v_y \right> \\
    \left< m v_z \right> \\
    \left< m v_x v_x \right> \\
    \left< m v_x v_y \right> \\
    \left< m v_x v_z \right> \\
    \left< m v_y v_y \right> \\
    \left< m v_y v_z \right> \\
    \left< m v_z v_z \right> \\
    \left< m v_x v^2 \right> \\
    \left< m v_y v^2 \right> \\
    \left< m v_z v^2 \right> \\
    \left< m v^4 \right> \\
  \end{pmatrix}
=
  \begin{pmatrix}
    \rho \\
    \rho u_x \\
    \rho u_y \\
    \rho u_z \\
    \rho u_x u_x + P_{xx} \\
    \rho u_x u_y + P_{xy} \\
    \rho u_x u_z + P_{xz} \\
    \rho u_y u_y + P_{yy} \\
    \rho u_y u_z + P_{yz} \\
    \rho u_z u_z + P_{zz} \\
    \rho u_x u^2 + \cdots + q_x \\
    \rho u_y u^2 + \cdots + q_y \\
    \rho u_z u^2 + \cdots + q_z \\
    \rho u^4 + \cdots + R_{iijj} \\
  \end{pmatrix}
  \, ,
\end{equation}

\noindent where some terms have been omitted for simplicity.
The full equations, including all terms and the convective fluxes are reported in Appendix~\ref{appendix:full-eqs}.
Some moments appearing in the fluxes $\bm{F}$ are unknown and can be found by solving the entropy maximisation problem:
\begin{itemize}
  \item First, one can find the coefficient vector, $\bm{\alpha}$, that corresponds to the state, $\bm{U}$, by numerical iterations. This gives the maximum-entropy distribution function, $f$, as expressed in Eq.~\eqref{eq:f14-definition}.
  \item Then, the distribution function can be integrated, with the proper weights, as to give the value of the required closing moments. 
\end{itemize}

As mentioned, this approach is very computationally demanding. 
Instead, we employ here a computationally affordable approximated solution of the entropy-maximization problem.
This formulation bypasses completely the computation of the coefficient vector, $\bm{\alpha}$, and 
gives the required closing moments as an algebraic function of the gas state, $\bm{U}$.
To date, this approach has been formulated for the 14-moment system,\cite{mcdonald2013affordable} and for its 
21-moment extension.\cite{giroux2021approximation}
As long as a direct knowledge of the shape of the distribution function (and thus the coefficient vector, $\bm{\alpha}$)
is not required during the solution, these approximations are sufficient for solving the moment system.

For an electric field with only components in the $(x,y)$ plane, the electrostatic source terms 
read\cite{boccelli2021moment}
\begin{equation}\label{eq:em-src}
  \begin{pmatrix}
    S_{\bm{E},1}  \\ 
    S_{\bm{E},2}  \\ 
    S_{\bm{E},3}  \\
    S_{\bm{E},4}  \\
    S_{\bm{E},5}  \\
    S_{\bm{E},6}  \\
    S_{\bm{E},7}  \\
    S_{\bm{E},8}  \\
    S_{\bm{E},9}  \\
    S_{\bm{E},10} \\
    S_{\bm{E},11} \\
    S_{\bm{E},12} \\
    S_{\bm{E},13} \\
    S_{\bm{E},14}
  \end{pmatrix}
  = 
  \frac{q}{m}
  \begin{pmatrix}
    0 \\                          
    E_x U_1 \\           
    E_y U_1 \\           
    0 \\
    2 E_x U_2 \\
    E_x U_3 + E_y U_2 \\
    E_x U_4 \\
    2 E_y U_3 \\
    E_y U_4  \\
    0 \\
    E_x \left( 3 U_5 + U_8 + U_{10} \right) + 2 E_y U_6 \\
    2 E_x U_6 + E_y \left( U_5 + 3 U_8 + U_{10} \right) \\
    2 E_x U_7 + 2 E_y U_9  \\
    4 E_x U_{11} + 4 E_y U_{12} 
  \end{pmatrix} \, ,
\end{equation}

\noindent where $U_i$ are the elements of $\bm{U}$.
The expression of the ionization source terms $\bm{S}_{\mathrm{iz}}$ depend on the specific test case and is given in the following sections.
During the numerical computations, one needs to know the system wave speeds for the sake of computing the Courant number and the numerical fluxes.
While it is possible to obtain the wave speeds numerically, by computing the eigenvalues of the flux Jacobian,
this increases significantly the computational times. 
Here, we have employed an approximated form of the wave speeds.\cite{baradaran2015development,boccelli2021moment}


While it describes well a wide range of non-equilibrium situations, the maximum-entropy method suffers from the impossibility 
to reproduce a set of otherwise physically acceptable states.
This issue was identified by Junk\cite{junk2002maximum} and appears as a singularity in the convective fluxes, in some regions of
moment space (denoted as ``Junk subspace'', or ``Junk line'' in 1D problems).

The value of the coefficient vector, $\bm{\alpha}$, cannot be determined for the states located on the Junk subspace, and the
value of some closing moments diverges.
For continuum collision-dominated problems, this appears not to be an issue, as the system does not depart excessively from equilibrium.
Moreover, the Junk subspace acts as to repel the system, that generally tends to circumvent it, instead of crossing it. 
However, when collisionless or rarefied gases are considered, one should pay particular numerical care.\cite{boccelli2021moment,boccelliInPreparation}
This happens to be the case for ions.
Indeed, in low-collisional circumstances, it is possible that the system develops two strongly non-equilibrium states that are located directly across 
this singularity.
It is also possible that numerical reconstruction procedures, employed in higher-order numerical schemes, predict states that are artificially
located on the Junk subspace.
In such scenarios, the numerical solution may become particularly stiff, or even diverge, and one may want to selectively employ first-order schemes
if this is the case.
Ultimately, this is highly test case dependent.
In the specific test cases of this work, the solutions appeared to be reasonably easily obtained.


\section{Test case I: axial acceleration}\label{sec:axial-accel}

Our first test case considers the one-dimensional axial acceleration of ions along the thruster centerline.
The geometry and plasma conditions for this test case have been previously studied in Boccelli et al,\cite{boccelli2020collisionless}
using different models (in that work, it is referred to as ``test case B'').
We wish to compare the maximum-entropy model against a kinetic solution, and we select the Particle-in-Cell (PIC) simulations 
of Boeuf \& Garrigues\cite{boeuf2018b} and Charoy et al.\cite{charoy20192d}
However, these PIC simulations are two-dimensional in the axial-azimuthal plane and are unsteady, due to the presence of travelling waves.
To reduce them to one dimension, we perform averages in the azimuthal direction and in time, resulting in a purely axial and steady state problem.

\begin{figure}[h!tb]
  \centering
  \includegraphics[width=1.0\columnwidth]{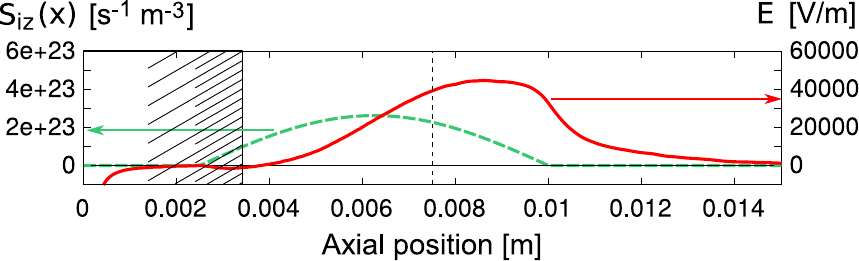}
  \caption{One-dimensional axial acceleration test case. Prescribed ionization profile (dashed green line) and electric field (solid red line). 
           The dashed domain is excluded from the computation. The vertical line at $x=0.0075$ corresponds to the maximum of the magnetic field in the PIC simulations 
           and can be interpreted as the channel exit plane.}
  \label{fig:one-dim-domain-E-S}
\end{figure}

The PIC simulations solve the fully coupled (electrons + ions) problem, while we are interested in studying the dynamics of ions alone.
Therefore, from the averages in the azimuthal direction and in time, we extract an averaged steady state electric field profile and impose it in our 
moment simulations. 
For simplicity and to speed-up convergence, the PIC simulations prescribed the ionization profile.
We do the same, and select it to be zero everywhere except in the region $x_1\le x \le x_2$, where
\begin{equation}\label{eq:prescribed-S-boeuf-cos}
  S_\mathrm{iz}(x) = S_0 \cos\left[\pi(x - x_M)/(x_2 - x_1)\right] \, ,
\end{equation}

\noindent with $x_1=0.0025$~m and $x_2=0.01$~m, $x_M=(x_1+x_2)/2$ and $S_0=2.62\times10^{23}~\si{m^{-3}s^{-1}}$.
Figure \ref{fig:one-dim-domain-E-S} shows the employed computational domain, the ionization source and the electric field.
Notice that, when seeking for a fluid one-dimensional steady-state solution with prescribed electric field and ionization profile, one risks to obtain a singularity in the ion density, at the point of ion velocity inversion.
Indeed, at the locations where the bulk velocity is low or zero, the locally produced cold ions can progressively cumulate,
resulting in a very localized maximum.
In such circumstances, thermal diffusion is the only process that can dissipate this peak, but such a process is slow due 
to the low ion birth temperature. 
This would not happen in a fully coupled scenario, 
where a local ion density peak would cause a strong restoring electric field.
Moreover, in real-life Hall thrusters, travelling waves and unsteadiness would introduce additional mixing effects.
Since this cannot be reproduced in the present simplified test case, we decided to avoid the said problems by 
artificially cropping the domain to the region of positive electric field, as shown in Fig.~\ref{fig:one-dim-domain-E-S}.
This choice still allows us to represent adequately the full acceleration region.

\subsubsection*{A lower-dimensionality model: the 1D 5-moment system}
In the present configuration, the evolution of ions only happens along the axial direction, as a result of the prescribed 
axial electric field.
Moreover, since all ion collisions are neglected, there is no momentum or energy relaxation among the three directions.
This case can be efficiently described by modelling the evolution of a single degree of freedom, 
thus considering a one-dimensional problem in both physical space and velocity (1D1V).
In such formulation, the particle velocity is thus represented by a scalar, $v$, and the moments are defined through
the following one-dimensional integrals:
\begin{equation}
  \begin{cases}
  \rho = \int m f(v) \, \mathrm{d} v \, , \\
  \rho u = \int m v f(v) \, \mathrm{d} v \, , \\
  P = \int m (v-u)^2 f(v) \, \mathrm{d} v \,  ,\\
  Q = \int m (v-u)^3 f(v) \, \mathrm{d} v \, , \\
  r = \int m (v-u)^4 f(v) \, \mathrm{d} v \, . \\
  \end{cases}
\end{equation}

The governing equations for such moments are obtained, as for the 14-moment case, by integrating the 
kinetic equation.
Ultimately, the 1D1V assumptions result in a system composed by 5 moments only, 
that still contains axial non-equilibrium but is much simpler to solve.
Notice that this system can also be obtained by simplifying the full 14-moment system,
setting to zero all velocities and pressure components, except for those in the considered direction. 
Out of the three velocity components, $u_{x}$, $u_y$ and $u_z$, only one is to be retained, and is denoted here by $u$.
The same happens to the heat flux vector.
In the pressure tensor, only the term $P_{xx} \equiv P$ is non-zero.
The 3D fourth-order moment, $R_{iijj}$, is the contraction of various indices of the tensor $R_{ijkl}$.
In the 1D1V case, there is no contraction to be performed, and this term is simply referred to as ``$r$''.
It should be stressed that that the definition of $Q$ employed here differs from the traditional definition of the heat flux, 
$q$, by a factor $1/2$.
Also notice that the value of the adiabatic coefficient for a 1D1V gas is $\gamma = 3$:
this consideration is important when comparing the Euler equations to the present 5-moment model.

The simpler 5-moment one-dimensional system reads\cite{mcdonald2013affordable}
\begin{equation}\label{eq:5mom-sys}
  \frac{\partial \bm{U}_5}{\partial t } + \frac{\partial \bm{F}_5}{\partial x} = \bm{S}_5^\mathrm{iz} + \bm{S}_5^{E} \, ,
\end{equation}

\noindent with state vector 
\begin{equation}\label{eq:U-5-definitions}
  \bm{U}_5
=
  \begin{pmatrix}
    \rho \\
    \rho u \\
    \rho u^2 + P \\
    \rho u^3 + 3 u P + Q \\
    \rho u^4 + 6 u^2 P + 4 u Q + r
  \end{pmatrix}
\end{equation}

\noindent and convective fluxes
\begin{equation}\label{eq:F-5-definitions}
  \bm{F}_5
=
  \begin{pmatrix}
    \rho u \\
    \rho u^2 + P \\
    \rho u^3 + 3 u P + Q \\
    \rho u^4 + 6 u^2 P + 4 u Q + r \\
    \rho u^5 + 10 u^3 P + 10 u^2 Q + 5 u r + s
  \end{pmatrix} \, .
\end{equation}

\noindent As discussed for the 14-moment case, 
the higher-order closing moments (here, $s$) can be closed by solving numerically 
the entropy-maximization problem and by further integrating the distribution function.
Here, we employ the interpolative closure of McDonald \& Torrilhon,\cite{mcdonald2013affordable} that approximates the
full entropy maximization procedure.

The ionization source terms are obtained by considering the moments for the neutral population and 
transforming them into ion quantities, at a rate given by $S_{\mathrm{iz}}(x)$, that 
is prescribed from Eq.~\eqref{eq:prescribed-S-boeuf-cos}.
The source terms read
\begin{equation}\label{eq:ch-ions-maxent-iz-source-5moments}
  \bm{S}_5^\mathrm{iz}
= \frac{S_\mathrm{iz}(x)}{n_n}
  \begin{pmatrix}
    \rho_n \\ \rho_n u_n \\ \rho_n u_n^2 + P_n \\ \rho_n u_n^3 + 3 u_n P_n  \\
    \rho_n u_n^4 + 6 u_n^2 P_n + r_n
  \end{pmatrix} \, ,
\end{equation}

\noindent where $\rho_n$ and $n_n$ are the neutral mass and number densities, $u_n$ and $P_n$ the average velocity and pressure of the
background neutrals, and with $r_n=3P_n^2/\rho_n$, and where we have assumed that the ion mass is approximately equal to 
the mass of neutrals. 
For consistency with the PIC test case, we take here $u_n = 0$.
The electrostatic sources read 
\begin{equation}\label{eq:electrostatic-source-5mom}
  \bm{S}_5^{E}
  =
  \frac{q}{m}
  \begin{pmatrix}
    0 \\ E \, U_1 \\ 2 \, E \, U_2 \\ 3 \, E \, U_3 \\ 4 \, E \, U_4 
  \end{pmatrix} \, .
\end{equation}

\noindent where $U_{1,\cdots,5}$ are taken from Eq.~\eqref{eq:U-5-definitions}.

\subsubsection*{Comparison of the results}

\begin{figure*}[h!tb]
  \centering
  \includegraphics[width=0.8\textwidth]{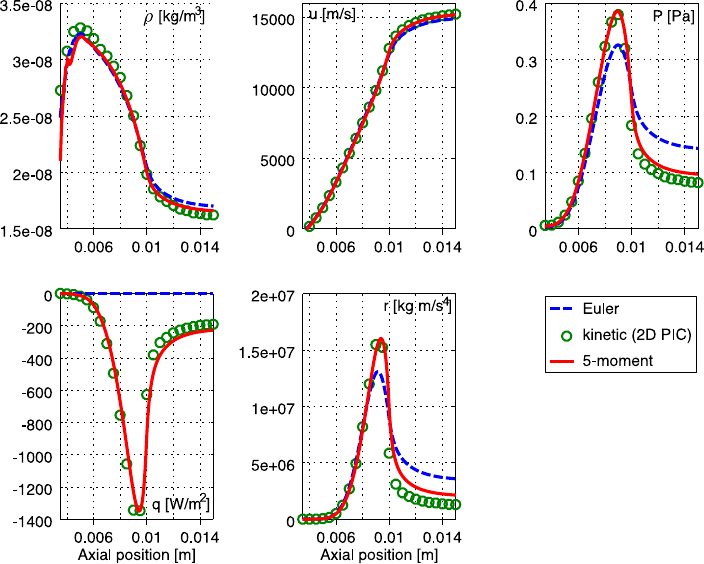}
  \caption{Evolution of axial quantities in a 1D Hall thruster channel, axial acceleration.
           A moment space plot is not shown, since the PIC solution embeds excessive statistical noise.}
  \label{fig:ch-ions-maxent-1D-hall-thruster}
\end{figure*}

In the simulations, xenon ions are considered, with mass $m=2.18\times10^{-25}~\si{kg}$.
Figure~\ref{fig:ch-ions-maxent-1D-hall-thruster} compares the solution of the maximum-entropy 5-moment system to the 
solution of the Euler equations and to the azimuthal and time averages of the PIC solutions of Charoy et al.\cite{charoy20192d}
First, one can notice that the lower order moments, density and bulk velocity, are accurately reproduced  
by the relatively simple Euler equations.
This is a known feature and results from the relatively large value of the average bulk velocity if compared to the ion temperature.
Indeed, as expected, ions are supersonic in the acceleration region and reach a Mach number $M\approx2$.
The strong convection hides the effects of non-Maxwellian distribution functions and the lower-order moments appear to be
practically unaffected by non-equilibrium.

On the other hand, the effect of non-equilibrium is much stronger on the pressure, heat flux and fourth-order moment.
This is because such moments are central (i.e., they are computed about the peculiar velocity $c = v - u$):
this removes convection from the picture, and the detailed shape of the distribution function plays an important role. 
For this reason, reproducing central moments is much more challenging, as the selected moment method must cope with the
task of recovering accurately the strong non-equilibrium states associated to the low collisionality.
This can be confirmed by looking at the VDFs for this problem.\cite{boccelli2020collisionless}
The Euler equations appear to under or over-estimate the pressure by roughly the 80\% in some regions, also due to 
the fact that the heat flux is neglected altogether in that model.
Besides the classical Euler variables (density, momentum and energy/pressure), 
Figure~\ref{fig:ch-ions-maxent-1D-hall-thruster} shows a fourth-order moment $r$ associated to the Euler equations.
This moment does not appear explicitly in the Euler system:
what we show here is the moment $r$ associated to a Maxwellian distribution at the pressure and density predicted by the Euler equations.
In other words, this is the equilibrium value of the fourth-order moment, given by $r = 3 P^2/\rho$ (in the present 1V case).
Together with the heat flux, the fourth-order moment can be employed as an effective measure of non-equilibrium, 
and it can have a direct effect on chemical reactions (although this is not considered in the present work).

The maximum-entropy method manages to reproduce the pressure, heat flux and fourth-order moment $r$ to a good 
accuracy.
Still, the results do not match exactly with the PIC simulations, and this can be due to some different factors.
First, the PIC results shown here include some statistical noise.
Also, the PIC profiles are obtained from azimuthal and temporal averages of unsteady simulations that would otherwise 
show azimuthally travelling waves.
In contrast, the maximum-entropy results are 1D and stationary.
However, these 2D and noise effects are known to play a very marginal role, as it was shown by Boccelli et al.,\cite{boccelli2020collisionless}
where they were compared to simplified steady state closed-form 1D kinetic solutions.
Therefore, it is the authors' belief that the main differences come from the assumptions of the maximum-entropy model itself.
Higher order maximum-entropy formulations are possible (7-moment system, in 1D1V)
and could improve the accuracy further, but this goes beyond the scope of the this work.

As mentioned, the maximum-entropy method is generally more expensive than the Euler equations.
However, for the present case, the problem appears to be rather simple:
in most of the domain, the solution does not cross the Junk line. 
There is some exception near the left boundary, around the artificial region of zero velocity, but this did not play any major role.
Since the solution stays rather far from the Junk line, the fluxes do not become singular and the system wave speeds do not diverge.
Ultimately, the maximum-entropy system is approximately 5--10 times more expensive than the Euler equations.
The only numerical difficulties may happen in the region where the ion bulk velocity reverses (axial position $x\approx0.004~\si{m}$).
For a further discussion of the computational costs associated to the maximum-entropy method, see Section~\ref{sec:computational-efficiency}.

\subsubsection*{Details on the numerical simulations}

The 5-moment and the Euler solutions are obtained with a finite-volume solver, on a grid of $1000$ cells, 
with second order accuracy in space (van Leer's MUSCL reconstruction and Rusanov fluxes 
with TVD van Albada symmetric limiter \cite{van1974towards,toro2013riemann,leveque2002finite}) and marching in time until convergence.
The implementation is based on the one-dimensional sub-version of the Hyper2D solver.\cite{hyper2Dgithub}


\section{Test case II: ion-wave trapping}\label{sec:ion-wave-trapping}

After studying the problem of the ion axial acceleration, we wish to consider another aspect of Hall thruster discharges, 
namely the capability of the maximum-entropy method to deal with azimuthal instabilities.\cite{boeuf2018b,taccogna2019numerical}
These instabilities result in travelling electric field waves, that cause the ion azimuthal VDF to show ion-wave trapping.
\cite{asadi2019numerical}
This phenomenon has been suggested to be responsible for the saturation of azimuthal electron drift instabilities. \cite{lafleur2016theory1}
Therefore, the capability to reproduce it with a fluid method such as the maximum-entropy method assumes a particularly interest.

As a test case, we consider a one-dimensional domain oriented along the azimuthal direction.
Any curvature is neglected for simplicity, but periodic boundary conditions are employed.
A sinusoidal electric field wave is imposed along the domain, with a potential
\begin{equation}
  V(y) = V_0 \cos(\omega t + k y) \, ,
\end{equation}

\noindent where the $y$ coordinate refers to the azimuthal direction, while $V_0$ is the amplitude and $\omega$ and $k$ are the angular 
frequency and wave number.
The electric field results from differentiation along $y$, and reads $E=V_0 k \sin(\omega t + k y)$.
The frequency and wave number are chosen as to provide reasonable results.
Considering a domain length $L=0.01~\si{m}$, the choice $k = 2 \pi N / L$, with $N=4$ results in four peaks inside the domain.
This is sufficient for our purposes of testing the method in presence of some periodicity.
A reasonable value for $V_0$ is here estimated by considering the simulations of Charoy et al,\cite{charoy20192d}
where the magnitude of the azimuthal electric field is approximately equal to $E_y \approx 2 \times 10^4~\si{V/m}$.
Considering the said choice for the wave number $k$, we thus select a value of $V_0 = 10~\si{V}$.

The effect of the electric field wave on the plasma ultimately depends on the average velocity of the ions (or the wave velocity)
and on the ratio of the electric potential, $V_0$, to the ion temperature, $T_i$.
If the ion bulk velocity is small, in the case of $V_0 > T_i\, [\si{eV}]$, the whole ion VDF can be expected to be strongly influenced
by the electric field. 
On the other hand, in the limit of $V_0 \ll T_i\, [\si{eV}]$, the ion possess much more thermal energy than the electric field, whose effect is 
little.
Different cases are expected to result in a different accuracy of the maximum-entropy model.
Here, we consider the case where $V_0 = 10~\si{V} \equiv T_i \, [\si{eV}]$, and thus initialize the ions at a temperature of $T_i = 10~\si{eV}$.
Finally, from the numerical dispersion relations,\cite{charoy20192d}
the waves appear to travel at a velocity $v=\omega/k$ that is roughly equal to 
twice of the ion thermal speed $v_{th}=\sqrt{8 k_B T_i/\pi m}$, with $k_B$ the Boltzmann constant.
This gives $\omega=21.745~\si{MHz}$.
These values define completely the problem, and, despite being rather arbitrary, they give a reasonable starting point for the sake 
of comparing the different methods.
For additional simulations, with different ratios of $V_0/T_i$, and for both moving and stationary waves,
the reader is referred to the thesis work by the first author.\cite{boccelli2021moment}

Along the domain, we initialize ion velocities from a Maxwellian VDF, at a temperature of $10~\si{eV}$ and with zero initial bulk velocity. 
Since collisionless conditions are considered, the particle velocity components $v_x,v_y$ and $v_z$ are decoupled, 
and there is no energy or momentum exchange among the three degrees of freedom.
Therefore, a maximum-entropy modelling of this problem is achieved here through the 5-moment system of Eq.~\eqref{eq:5mom-sys}.
No ionization sources need to be considered, since the domain is periodic and the total mass is thus conserved.

\begin{figure*}[h!tb]
  \centering
  \includegraphics[width=\textwidth]{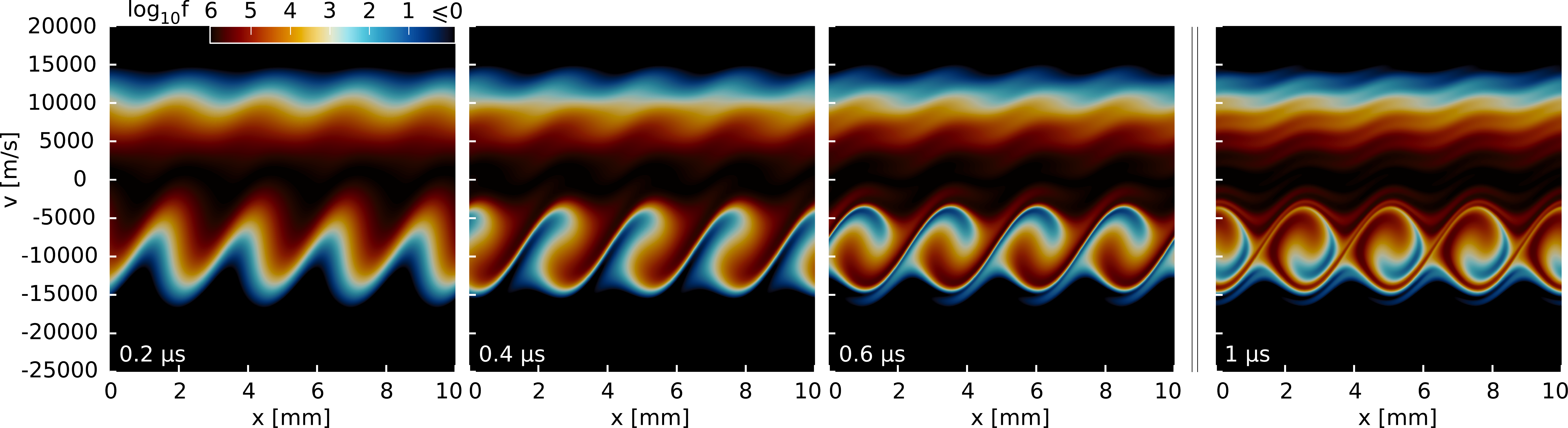}
  \caption{Travelling electric field test case. Finite-volume solution of the 1D1V kinetic equation for the velocity distribution function $f$. Logarithmic scaling.}
  \label{fig:kin-sol-ions-moving-wave}
\end{figure*}

\begin{figure*}[h!tb]
  \centering
  \includegraphics[width=0.75\textwidth]{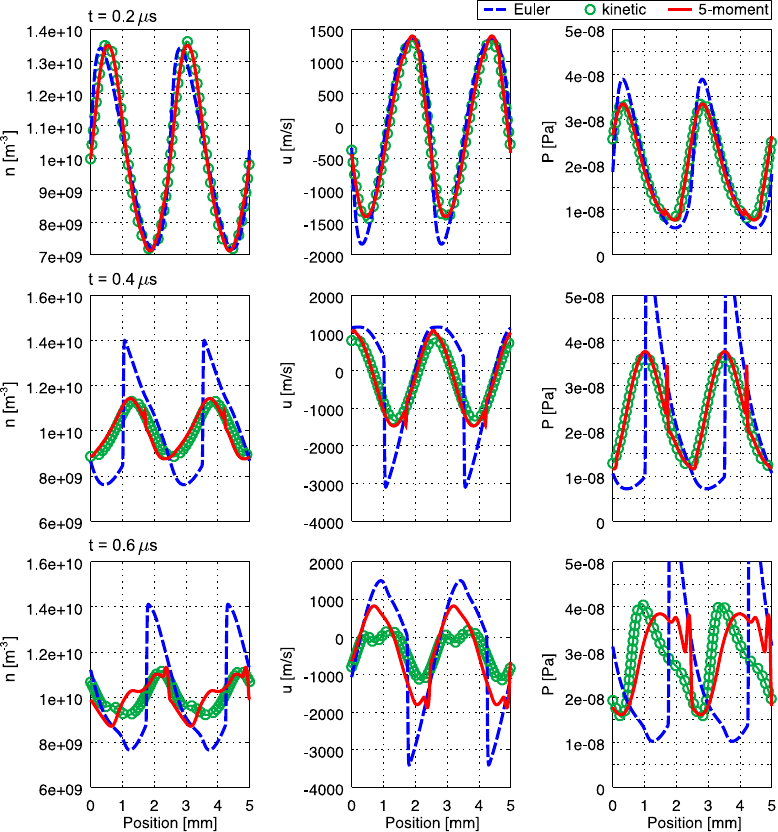}
  \caption{Travelling electric field test case. Density, velocity and pressure at three time steps. Symbols: kinetic solution. Red line: 5-moment system. Blue dashed line: Euler system.}
  \label{fig:moments-moving-wave-rho-u-P}
\end{figure*}

\begin{figure*}[h!tb]
  \centering
  \includegraphics[width=0.75\textwidth]{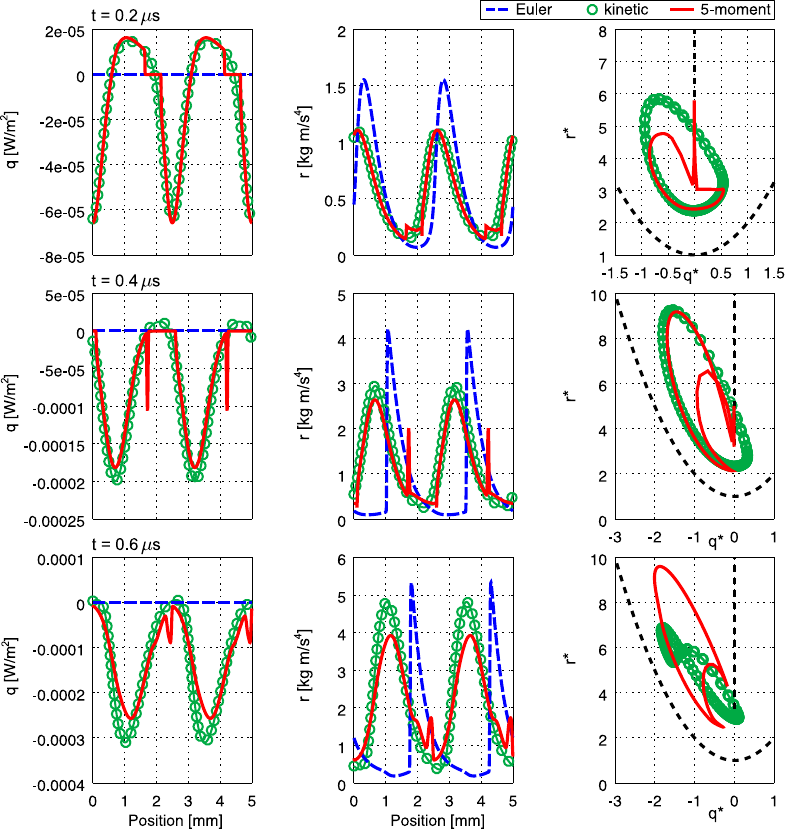}
  \caption{Travelling electric field test case. Heat flux and fourth-order moment at three time steps. Symbols: kinetic solution. Red line: 5-moment system. Blue dashed line: Euler system. The black dashed parabola and vertical line in the Right-column plots represent the physical realizability boundary and the Junk subspace respectively.}
  \label{fig:moments-moving-wave-q-r-qr}
\end{figure*}

First, Eq.~\eqref{eq:kinetic-eq} is rewritten for a particle with a single translational degree of freedom (``1D1V'' phase space), and neglecting the collision operator. 
The domain is discretized in phase space using a two-dimensional grid (one dimension for physical space $x$ and one for the velocity axis $v$) 
with periodic conditions along $x$ and solved with a finite-volume method based on the scheme of Mieussens. \cite{mieussens2000discrete}
Additional details are given below. 
The computational domain is shown in Figure~\ref{fig:kin-sol-ions-moving-wave}, that also shows the time evolution of the initial Maxwellian:
the negative-velocity side of the VDF is affected by the travelling electric field wave and curls in phase-space showing 
ion-wave trapping.
In the absence of collisions, this effect keeps increasing, the VDF creates whorls and spikes in phase space
and we expect the solution to be progressively harder to be followed by a fluid method.
With the employed grid, composed of $2000\times2000$ cells in phase space, some numerical diffusion starts to be slightly visible in the phase space plots after a time of $0.5-1~\si{\mu s}$ due to the sharp
gradients inside the trapped ions loops.
However, this has no appreciable effect on the five moments of interest: density, velocity, pressure, heat flux and fourth-order moment, $r$.
For each position, $x$, the moments of the kinetic solution are computed by numerical integration of the VDF along the velocity axis.
A simple first order integration scheme is employed and no significant change is observed by using finer or coarser grids.

The moments are then compared to a solution of the 5-moment and to the Euler systems.
The results are shown in Fig.~\ref{fig:moments-moving-wave-rho-u-P} for the density, velocity and pressure fields, 
and in Fig.~\ref{fig:moments-moving-wave-q-r-qr} for the heat flux and fourth-order moment.
The Euler equations appear to lose accuracy quite early in the simulation, introducing unexpected shock waves and deviating from the kinetic solution.
On the other hand, the maximum-entropy model manages to follow the kinetic solution with a good accuracy in the first stages of the 
simulation, eventually degrading in accuracy as time passes.
The presence of some artificial peaks can be observed in the maximum-entropy solution, and it is most likely associated to the crossing
of the Junk line, as can be seen in Fig.~\ref{fig:moments-moving-wave-q-r-qr}-Right 
(for the 5-moment system, the Junk subspace is located along the vertical line $q^\star=0, r^\star> 3$ in dimensionless moment space, where 
nondimensionalization is performed following Eq.~\eqref{eq:dimensionless-moments-def}).

This test case confirms the increased accuracy of the maximum-entropy approach, but also shows its limitations.
As non-equilibrium develops, more and more moments would be needed to obtain an accurate solution.
It should be stressed that this test case is particularly challenging, since the progressive build-up of VDF whorls in phase space
causes a constantly increasing non -equilibrium.
The present one-dimensional simulation with periodic conditions forces the ions to remain in these conditions.
On the other hand, in a real scenario, ion-wave trapping happens only in a certain part of the domain:
ions transit through this region for a limited amount of time, and are eventually accelerated outwards.
In a real scenario, we should thus expect to encounter somehow intermediate non-equilibrium, as we shall study in Section~\ref{sec:full-test-case}.

\subsubsection*{Details on the numerical simulations}

The fluid simulations (Euler and maximum-entropy system) are obtained using the one-dimensional sub-version of the Hyper2D solver,\cite{hyper2Dgithub} after implementing the relevant equations and source terms. 
The 1D1V kinetic equation is obtained using the kinetic theory sub-version of Hyper2D.

The kinetic simulations are computed on a phase-space grid composed of $2000\times2000$ cells,
Second order upwinded numerical fluxes are employed\cite{mieussens2000discrete} and the solution is advanced in time with a 
second order explicit midpoint Euler scheme.
The time step is chosen as to give a Courant number of $0.5$, based on the advection in physical space 
($v_{\mathrm{max}}= 25\,000~\si{m/s}$ for the employed domain) and in velocity space 
(where the maximum acceleration is $q\,\mathsf{E_{\mathrm{max}}}/m$).

Fluid simulations are computed on a grid of $2000$ cells, using second order accuracy in space (Rusanov numerical fluxes, 
van Leer's MUSCL linear reconstruction, with symmetric van Albada limiter) and a midpoint Euler second order explicit time marching scheme.
The maximum-entropy simulations require to set a limiting parameter $\sigma_{\mathrm{LIM}}$ for computing the closing moments:
the present test case appears to be quite sensitive to that parameter, and we have here employed a value of $\sigma_{\mathrm{LIM}} = 10^{-5}$ 
(for all details, see McDonald \& Torrilhon\cite{mcdonald2013affordable} and see the further discussion in Section~\ref{sec:computational-efficiency}).


\subsection{Analysis of the computational cost}\label{sec:computational-efficiency}

This test case proves ideal for illustrating the computational cost associated to the fourth-order maximum-entropy method.
We sketch here a simplified analysis of the numerical cost $C$ of our finite-volume computations, expressed in terms of 
total number of operations, and excluding memory access issues.
Considering an explicit time marching scheme, we can write
\begin{equation}\label{eq:cost-of-a-solution}
  C \propto a \times N^d \times N \times \lambda_{\mathrm{max}} \, .
\end{equation}

\noindent Here:

\begin{itemize}
  \item $a$ depends on the complexity of the PDEs;
  \item $N^d$ is the total number of simulated cells ($d$ is the dimensionality of the grid, and we assume $N_x = N_y = N$ for simplicity);
  \item the term $N \times \lambda_{\mathrm{max}}$ arises from the CFL constraint, that imposes a maximum time step that depends on the grid size and on 
        the maximum wave speed $\lambda_{\mathrm{max}}$.
\end{itemize}

\noindent The effect of source terms is neglected in the previous equation, as it does not play an important role in the present test cases.
From Eq.~\eqref{eq:cost-of-a-solution}, the overall cost of a deterministic Boltzmann simulation is seen to be proportional to $N^3$ in 1D1V, where $d=2$
(or $N^7$ for 3D3V simulations, where $d=6$), while the cost of a fluid simulation scales
as $N^2$ in 1D (or $N^4$ in 3D).
On the other hand, the parameter $a$ happens to be very small for the Boltzmann equation (only one scalar quantity $f$ appears in each cell), 
a bit higher for the Euler equations (three scalar quantities per cell in 1D, or five in 3D, and the need to compute primitive variables) 
and even higher for the maximum-entropy system (5 or 14 quantities per cell, plus the need to 
compute closing fluxes through a number of operations).
However, in the present test cases, the most noteworthy role is played by $\lambda_{\mathrm{max}}$.
In the Boltzmann solution, $\lambda_{\mathrm{max}}$ is uniquely determined by the velocity grid and the maximum applied electric field.
In the Euler equations, $\lambda_{\mathrm{max}}$ depends on the gas velocity and temperature.
For the maximum-entropy simulations, $\lambda_{\mathrm{max}}$ still depends on velocity and temperature, but also on higher order moments. 
In particular, if the solution approaches the Junk subspace (line $q^\star=0, r^\star>3$ in moment space), fluxes become singular
and the wave speeds may increase by several orders of magnitude. 
This drastically affects the computational cost.
Providing an a priori estimate for this effect is not trivial and is to be evaluated on a case-by-case basis.
For instance, from Fig.~\ref{fig:moments-moving-wave-q-r-qr}-Right, one can see that the solution approaches the Junk line at the initial stages of the simulation,
but then recovers completely after roughly $0.4~\si{\mu s}$.
The crossing of the Junk subspace is associated to a jump in moment space and to spikes in the spacial solution.

\begin{figure}[h!tb]
  \centering
  \includegraphics[width=\columnwidth]{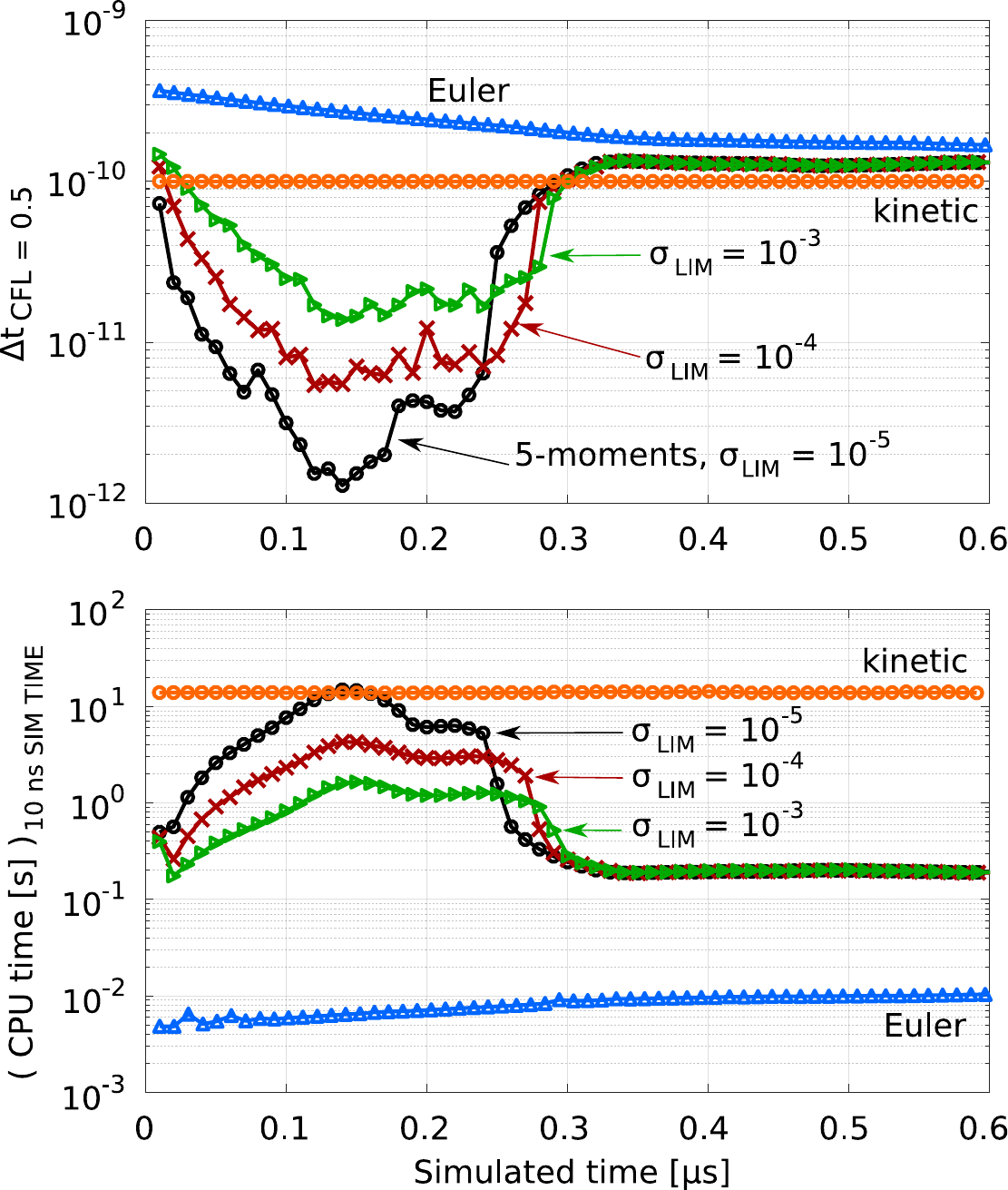}
  \caption{Comparison of different solvers. Top: maximum allowable time step for a fixed Courant number of $0.5$. Bottom: CPU time necessary to simulate $100~\si{ps}$.}
  \label{fig:comp-efficiency-time-step}
\end{figure}

In Fig.~\ref{fig:comp-efficiency-time-step} we compare the computational efficiency associated to the different methods, for the simulations of test case II.
The comparison is carried as the simulation evolves, since $\lambda_{\mathrm{max}}$ changes during the simulation.
For all test cases, we employ a grid of $2000$ cells (for the kinetic simulation, $2000\times2000$ cells)
and all test cases are performed on a single core.
Figure~\ref{fig:comp-efficiency-time-step}-Top shows the time step $\Delta t$ that guarantees a Courant number of $0.5$.
This can be seen to be constant throughout the simulation for the kinetic solver (since we employ fixed velocity grid and fixed forces).
For the Euler equations, the maximum time step decreases roughly two times during the simulation, due to an increase in temperature and/or velocity.
The maximum-entropy equations show the strongest limitations on $\Delta t$, due to the proximity of the solution to the Junk subspace, resulting in large wave speeds.
Figure~\ref{fig:comp-efficiency-time-step}-Bottom shows the CPU time required to advance the simulation by $10~\si{ns}$.
This is a measure of the overall computational cost, including all effects in Eq~\eqref{eq:cost-of-a-solution}.
For this benchmarking, the solver was run on an Intel\textsuperscript{\textregistered} Core\texttrademark ~i5-5300U CPU.
The severe time step limitation of the maximum-entropy method is balanced by a the lower grid dimensionality, and the method appears to be
always equally expensive or cheaper than the direct kinetic solution.
For the present 1D1V plasma problems, maximum-entropy simulations are roughly ten times heavier than Euler simulations, 
and somewhere between 0 and 100 times cheaper than direct kinetic simulations.
However, this strongly depends on the number of cells and on the dimensionality of the problem.
In 1D1V, for a sufficiently large number of cells, we expect the maximum-entropy simulations to 
maintain an analogous efficiency if compared to the Euler equations (both scale with $N^2$), 
but to improve its efficiency even further with respect to the direct kinetic method (that scales as $N^3$).

In Figure~\ref{fig:comp-efficiency-time-step}, three different simulations are shown for the maximum-entropy method, for different values of the 
parameter $\sigma_{\mathrm{LIM}}$, that appears in the expression of the closing moments.\cite{mcdonald2013affordable}
Typical suggested values\cite{boccelli2021moment,boccelliInPreparation} are in the range of $\sigma_{\mathrm{LIM}}\in[10^{-5}, 10^{-4}]$.


\section{Test case III: axial-azimuthal plane}\label{sec:full-test-case}

In this test case, we combine the axial acceleration problem of Section~\ref{sec:axial-accel} to the ion-trapping problem of Section~\ref{sec:ion-wave-trapping}. 
A two-dimensional domain is studied, with dimensions $L_x = 0.025~\si{m}$ and $L_y=0.0128~\si{m}$, following the simulations of Charoy et al.\cite{charoy20192d}
Periodicity is imposed along $y$.
We manufacture and prescribe an electric field, composed of an axial accelerating component, $E_x$, and an azimuthally travelling wave, $E_y$.
\begin{equation}
  \begin{cases}
    \bm{E}(x,y,t)   &= {E}_x(x) \,  \bm{\hat{x}} + {E}_y(x,y,t) \, \bm{\hat{y}} \, ,\\
    {E}_x(x)     &= {E}_0 \exp\left[ - (x-x_0)^2/L_0^2 \right] \, ,\\
    {E}_y(x,y,t) &= \alpha {E}_x(x) \sin(\omega_y t + k_y y) \, .
  \end{cases}
\end{equation}

\begin{figure}[h!tb]
  \centering
  \includegraphics[width=\columnwidth]{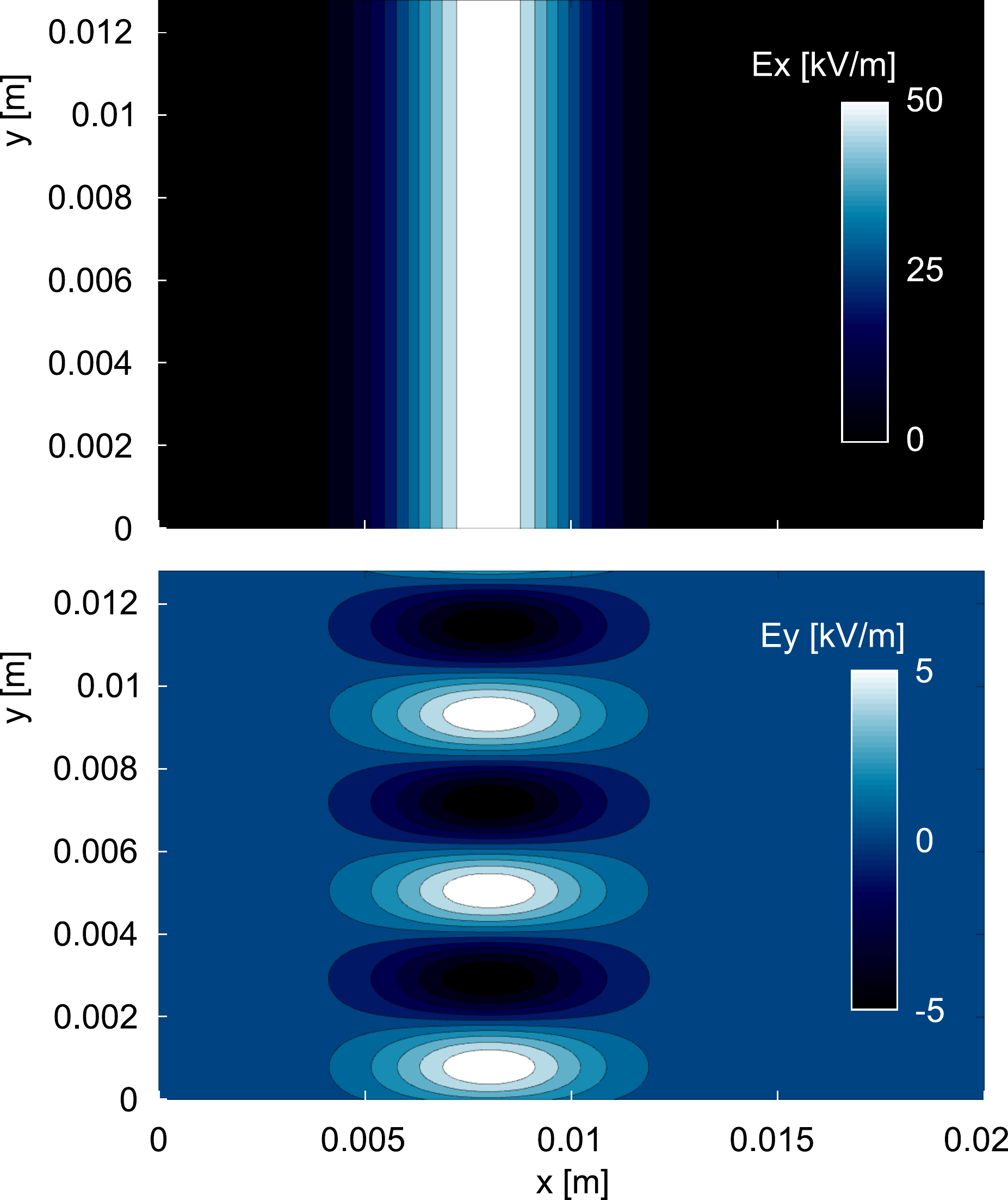}
  \caption{Imposed electric field at time $t=0$~s. The $E_y$ wave moves towards negative values of $y$.}
  \label{fig:ch-ions-maxent-2D-case-Ex-Ey}
\end{figure}

\noindent The axial field is shaped as a Gaussian centered at $x_0=0.008~\si{m}$, with a width $L_0=0.0025~\si{m}$ and a maximum amplitude of $E_0=50\,000~\si{V/m}$.
The azimuthal component, $E_y$, has an amplitude that is reduced with respect to $E_x$ by a factor $\alpha = 0.1$.
The value of $k_y$ is chosen here as to result in three peaks inside the domain, $k_y=6\pi/L_y$. 
This is not entirely coherent with the original simulations, where the electric field dynamics is observed to happen at a smaller spatial 
scale (roughly, 10 times smaller).
Given the preliminary nature of this work, we preferred to relax the spatial scale, as to allow for coarser numerical grids. 
Yet, it is important to preserve the correct azimuthal wave velocity, $v = \omega/k$. 
Therefore, the angular frequency is chosen as $\omega_y=2~\si{MHz}$. 
The electric field at a given time step is shown in Fig.~\ref{fig:ch-ions-maxent-2D-case-Ex-Ey}.

As done in Section~\ref{sec:axial-accel}, the ionization profile is imposed.
Its value is zero everywhere except in the region $x_1\le x \le x_2$, where
\begin{equation}\label{eq:prescribed-S-boeuf-2}
  S_\mathrm{iz}(x) = S_0 \cos\left[\pi(x - x_M)/(x_2 - x_1)\right] \, ,
\end{equation}

\noindent where $x_1=0.0025~\si{m}$, $x_1=0.01~\si{m}$, $x_M=(x_2+x_1)/2$ and with $S_0=6.62\times10^{23}$ $\si{[m^{-3}s^{-1}]}$.
This choice of $S_0$ results in a maximum ion current density of $J_{\mathrm{max}}=200~\si{A/m^2}$.
These values are taken from the PIC test cases\cite{boeuf2018b,charoy20192d} and are rather low for a real thruster.
However, this does not impact the present simulations, since there is no coupling with the electrons.

\begin{figure*}[h!tb]
  \centering
  \includegraphics[width=0.9\textwidth]{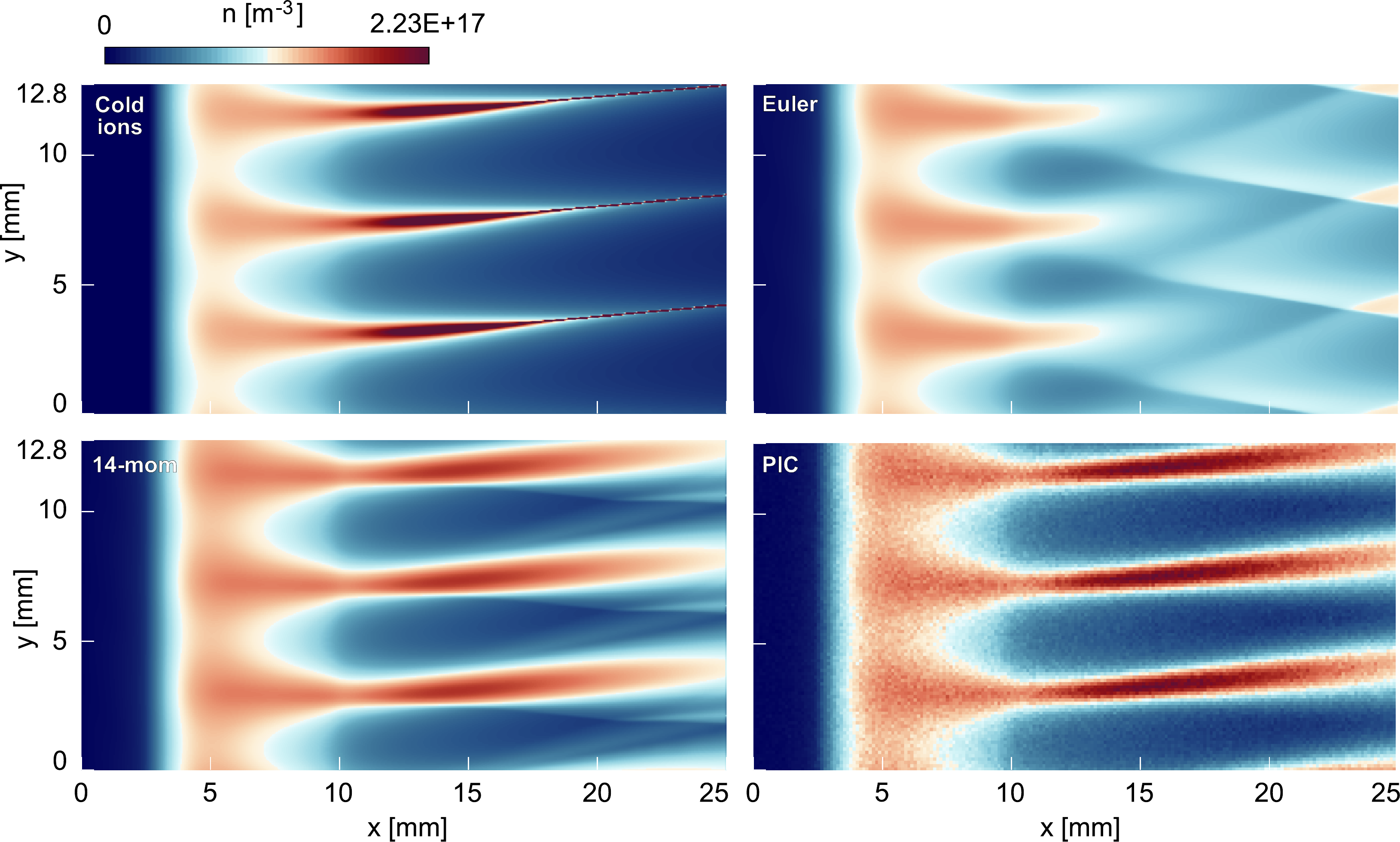}
  \caption{2D ion evolution inside an axial and a travelling azimuthal electric fields. Number density at time $t=50~\si{\mu s}$ as predicted by different models.}
  \label{fig:ch-ions-maxent-2d-results-density}
\end{figure*}

\begin{figure*}[h!tb]
  \centering
  \includegraphics[width=0.9\textwidth]{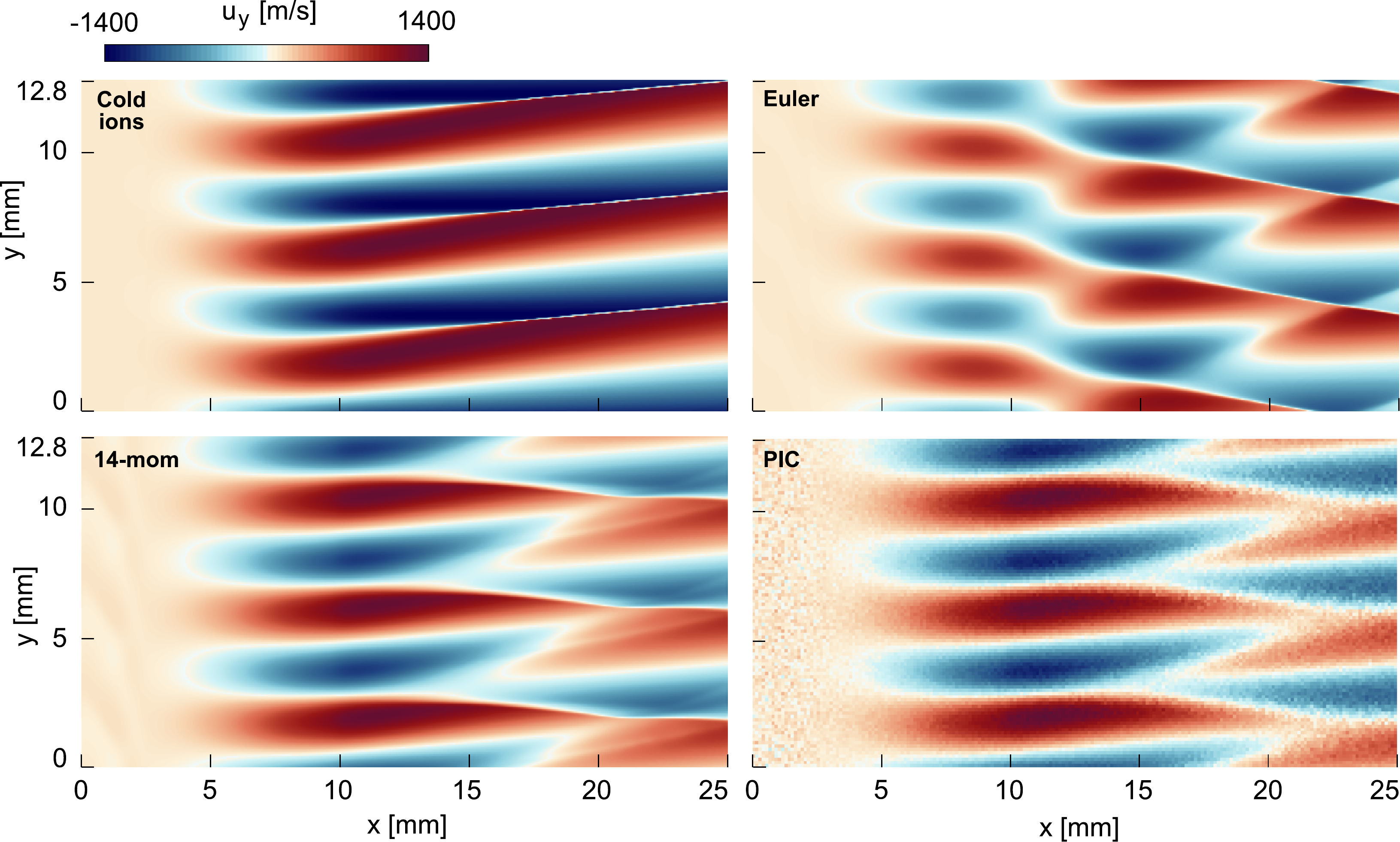}
  \caption{2D ion evolution inside an axial and a travelling azimuthal electric fields. Azimuthal velocity $u_y$ at time $t=50~\si{\mu s}$ as predicted by different models.}
  \label{fig:ch-ions-maxent-2d-results-velocity}
\end{figure*}

Four different modelling strategies are compared for this test case.
First, we test a simple fluid approach based on the pressureless gas equations. \cite{chen2003formation,bouchut2003numerical}
This simple approach is often employed in both fully fluid and hybrid kinetic-ion/fluid-electron simulations.
A solution of the Euler equations is also proposed, together with a solution of the 14-moment maximum-entropy system.
Finally, a kinetic solution is obtained by use of a simple particle method (Particle-in-Cell with prescribed electric field).
In all four simulations, only ions are considered and no collision is included.
After an initial transitory, the solutions soon settle to a steady state.
The four solutions are compared in Figures \ref{fig:ch-ions-maxent-2d-results-density} and \ref{fig:ch-ions-maxent-2d-results-velocity}
for the density and azimuthal velocity fields respectively.
The pressureless gas simulations appear to reproduce a reasonable density field, although, being a non-strictly hyperbolic system, 
they reproduce the characteristic delta-shocks.\cite{tan1994delta}
The Euler equations do not increase much the accuracy, but instead predict a peculiar shock structure.
On theoretical grounds, the Euler equations assume a Maxwellian VDF and thus full collisionality, while the present situation is fully 
collisionless. 
Most shocks are therefore unphysical and we shall rather have continuous smooth profiles.

\begin{figure*}[h!tb]
  \centering
  \includegraphics[width=0.9\textwidth]{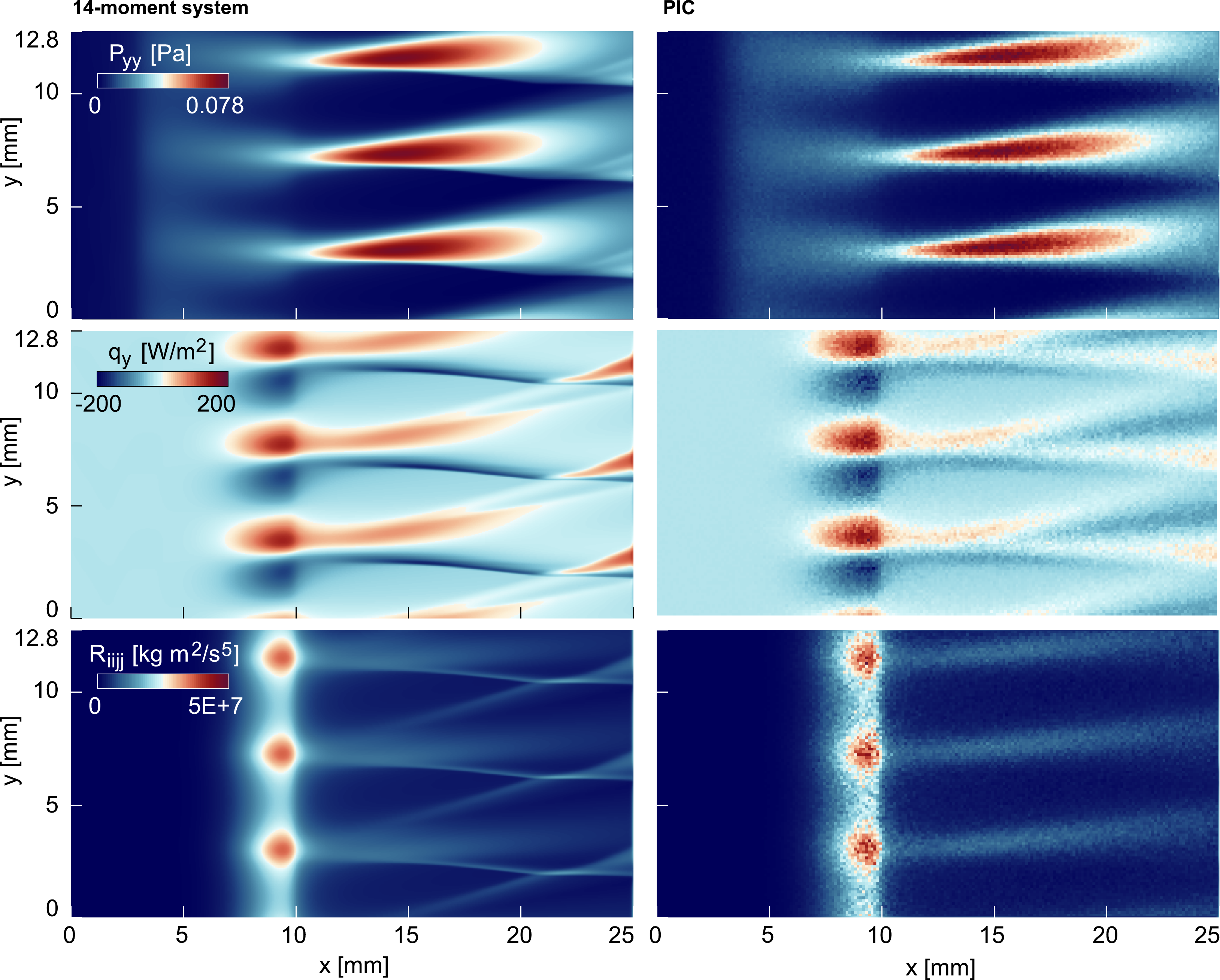}
  \caption{2D ion evolution inside an axial and a travelling azimuthal electric fields. Comparison of the 14-moment system and PIC solutions for selected moments. Time $t=50~\si{\mu s}$.}
  \label{fig:ch-ions-maxent-2d-results-14mom-PIC-somemoments}
\end{figure*}

The 14-moment system on the other hand manages to reproduce the kinetic solution to a good accuracy.
Some unphysical shocks are visible in the background, but appear as a small perturbation to an otherwise accurate bulk solution.
The accuracy of the 14-moment method is confirmed by Fig.~\ref{fig:ch-ions-maxent-2d-results-14mom-PIC-somemoments}, where additional moments
are compared to the kinetic ones.
Again, except for the presence of some additional waves with a rather small amplitude, the 14-moment system appears to reproduce
the kinetic solution to a good accuracy.
As anticipated in Section~\ref{sec:ion-wave-trapping}, in the present two-dimensional case the maximum-entropy approach shows a 
very high accuracy, as compared to the previous one-dimensional azimuthal simulations.
Indeed, in the 2D scenario, ions can escape from the azimuthal wave region due to the concurrent axial acceleration.
A strong non-equilibrium is realized, but the 14-moment model shows able to manage it.

We believe the solution to be substantially independent from the chosen size of the computational domain. 
Indeed, the azimuthal periodicity is enforced not only by the periodic boundary conditions, but also by 
imposing the electric field profile.
We do not expect that adding any integer number of wavelengths would modify the results.
On the other hand, we believe that parametric studies of the accuracy of the 14-moment method 
at different wavelengths are interesting, and are suggested as a future work.
Regarding the axial direction, one should consider that ions are supersonic. 
Therefore, we expect little to no effect on the simulated region, if one was to include a longer plume.

\subsubsection*{Details on the numerical simulations}

The kinetic solution is obtained with a simplified version of the Pantera Particle-in-Cell solver\cite{boccelli202014,parodi2021pic}
with a first-order explicit time marching scheme.
Since the electric field is imposed and there are no collisions, each simulated particle is independent.
As a result, the number of simulated particles plays no role over the final results, and only affects the noise of the output moments.
For the same reason, the space grid plays no role in the results and only defines the space resolution of moments.
The simulated particles are pushed with an explicit first order forward Euler integrator.
A time step $\mathrm{d}t = 10^{-8}~\si{s}$ is employed. 
Lower time steps did not show any appreciable difference in the resulting moments.

All fluid solutions (pressureless gas, Euler and 14-moment system) are obtained with the 
Hyper2D CUDA-accelerated finite-volume solver, \cite{hyper2Dgithub} modified as to solve the relevant sets of equations 
and source terms, and run on NVIDIA GPUs (Tesla K20X and/or A100 40GB). 
All solutions are obtained with double precision. 
While this is optional for the pressureless gas and for the Euler equations, this is a hard requirement for 
the 14-moment system. 
This is probably due to bad conditioning of the matrices employed in the interpolated closure, for certain states.

The numerical discretization employed $640\times320$ cells, 
with second order in space (van Leer's MUSCL linear reconstruction with symmetric van Albada slope limiter) and Rusanov fluxes. 
In the maximum-entropy simulations, reaching second order in this collisionless scenario proves to be rather difficult and the discretized 
system is prone to diverging.
To alleviate this, we modify the slope limiter, multiplying it by a factor $\beta = 0.7$. 
This has the effect of reducing the slope used in the linear reconstruction.
As a result, the accuracy goes below the second order, but the procedure still reduces significantly the otherwise large dissipation of 
the Rusanov flux alone.

%


A second-order Midpoint Euler time integration scheme is employed in all fluid simulations.
The time step, $\mathrm{d} t$, is chosen as to result in a maximum Courant number of $0.5$, throughout the domain. 
The time step is further limited during the simulation, such that $\mathrm{d} t \le 10^{-9}$~s.
This helped to deal with electrical and ionization source terms, that could be more demanding than the CFL
constraint, in certain conditions. 
In particular, this additional constraint proved to be particularly important at the beginning of the simulation, 
where the domain is initialized as empty. 
The simulations are repeated with different time steps, 
confirming that the time integration error is negligible.


For the present test case, providing a meaningful comparison between the computational cost of the fluid methods and that of the 
particle methods is not trivial.
Indeed, one should consider the following points.
First of all, in the present test case, all particles are independent from each other since the electric field is imposed
and since all collisions are neglected.
Therefore, there is no lower constraint on the number of particles to be employed in the present PIC simulations, 
and the only effect of such a choice lies in the accepted level of noise in the output moments. 
Moreover, even more importantly, a significant cost of PIC methods is represented by the particle sorting algorithm 
(particle-to-grid mapping), necessary for the computation of the electromagnetic fields.
This step appears to be the leading cost in certain parallel computations\cite{claustre2013particle}
but this cannot be analyzed in the present simplified test case.
For this reason, we defer the comparison of the fluid and particle efficiencies to future studies and 
refer the reader instead to the estimates of Section~\ref{sec:computational-efficiency}.

\section{Self-consistent plasma simulations}\label{sec:self-consistent}

In the previous sections, the coupling of ions with other charges has been avoided by prescribing 
instead reasonable values for the electric field.
In this section, we briefly discuss how one can embed the presented model into self-consistent 
simulations.
Only some very simple examples will be discussed here, and we leave the
discussion of full self-consistent simulations of $\bm{E}\times\bm{B}$ 
discharges to future works.


\subsection{Quasi-neutral model}\label{sec:quasi-neutral-model}

The simplest coupling between ions and electrons is probably achieved through the 
quasi-neutrality assumption,
where the electron number density and average velocity are $n_e = n_i \equiv n$ and $\bm{u}_e = \bm{u}_i$.
The electric field appearing in the source terms of the moment system, 
Eqs.~\eqref{eq:em-src} and \eqref{eq:electrostatic-source-5mom}, 
can be expressed by use of the electron momentum equation:
neglecting the electron inertia, one has
\begin{equation}
  e n \bm{E} = - \bm{\nabla} P_e \, ,
\end{equation}

\noindent where $P_e = n k_B T_e$ is the electron pressure.
For the one-dimensional 5-moment system, this gives 
\begin{equation}\label{eq:quasi-neutral-sources-5mom}
  \left(\bm{S}_5^{E}\right)_\mathrm{Q.N.}
=
- Z
  \begin{pmatrix}
    0 \\  \, 1 \\ 2 \, U_2/U_1 \\ 3\, U_3/U_1 \\ 4 \, U_4/U_1
  \end{pmatrix}  
  \frac{\partial P_e}{\partial x} \, ,
\end{equation}

\noindent where $Z=q/e$ is the ion charge number. 
In the previous equation, all unlabeled quantities refer to ions, 
and we have $U_1 \equiv \rho$, $U_2 \equiv \rho u^2$ etc, from Eq.~\eqref{eq:U-5-definitions}.
All the other entries in Eq.~\eqref{eq:5mom-sys} for the ion dynamics are unchanged.
Source terms for the 14-moment system are obtained analogously and involve the derivatives of 
$P_e$ in the $x$ and $y$ directions. 
If one assumes that the electron temperature $T_e$ is constant and uniform, then the full set of equations 
is closed.
Otherwise, one can supplement the model with an equation for the electron energy, 
depending on the specific needs of the problem to be studied.\cite{mora2005collisionless}
The quasi-neutral model introduced in this section is further discussed in 
Appendix \ref{app:examples-self-consistent}, where we consider the results of a numerical computation.


\subsection{Multi-fluid, hybrid simulations and overall computational cost}

Another possibility to obtain a self-consistent plasma model consists in employing 
the maximum-entropy system in a multi-fluid framework, or by coupling it to a stochastic particle 
description for one species.
In such cases, the density of each species is known throughout the simulated domain at each time step, 
and can be used to obtain the electric field by solving the Poisson equation, $\nabla \cdot \bm{E}=(q_i n_i - e n_e)/\varepsilon_0$,
or by solving the full set of Maxwell equations.
This approach allows one to describe situations of strong charge unbalance, such as plasma
sheaths,\cite{alvarez2020plasma} appearing for instance in the anode region and near the walls of Hall thrusters.

Coupling the ions with evolution equations for the electrons would introduce further 
time scales into the problem, associated to the Debye length and the electron dynamics. \cite{laguna2020asymptotic}
This typically results in a tighter time step limit for numerical simulations 
(consider for instance the electron and ion plasma frequencies, $\omega_e \gg \omega_i$).
If an explicit time marching numerical scheme is employed, for each ion time step, 
a large number of (smaller) electron time steps need to be performed.
Therefore, the electron module is often the most time-consuming part of a simulation. 
This partially mitigates the larger computational cost of the maximum-entropy ion model, if 
compared to the cheaper classical fluid models:
although the ion module becomes more expensive, this may be negligible if compared to the
electron module, overall.
An accurate estimation of the overall efficiency, as well as a comparison with a kinetic ion model, 
depends on the specific problem under consideration:
as discussed in Section~\ref{sec:computational-efficiency}, the cost of the maximum-entropy method
strongly depends on whether the solution approaches or not the Junk singularity.


\section{Conclusions}

This work considers the application of the 14-moment maximum-entropy system to the modelling of collisionless ions in Hall thruster discharges.
This system is the simplest fourth-order member of the maximum-entropy family of moment methods.
All 14 moments are employed to describe ions.
Electrons and neutrals are not explicitly simulated, and the electric field is instead prescribed: 
the simulations discussed in this work are intentionally decoupled.
This choice allows us to obtain a direct comparison among different ion models.
Some guidelines for performing self-consistent simulations are also given and an example of a quasi-neutral simulation is briefly discussed.

Three different test cases are analyzed.
First, the study of the production and axial acceleration of ions along a one-dimensional thruster channel and in the near plume is considered.
The maximum-entropy method appeared to reproduce the kinetic results to a good accuracy, bringing significant improvements over the 
simpler Euler equations of gas dynamics.

Then, a 1D azimuthal problem is considered, aimed at reproducing the azimuthal behavior of ions in the presence of ion-wave trapping.
This problem shows a progressively increasing degree of non-equilibrium.
Again, the maximum-entropy method appeared as a higher-accuracy alternative to the Euler equations, although, eventually,
also the maximum-entropy system departs from the kinetic solution, once a strong enough non-equilibrium is reached.

Finally, the ion evolution in the two-dimensional axial-azimuthal plane is considered.
Such test case combines the axial acceleration and the azimuthal ion-wave trapping test cases.
In this case, we also consider the often employed pressureless gas (cold ion) model.
The 14-moment system showed able to reproduce the kinetic solution to a high accuracy.
The presence of an axial acceleration field allowed to limit the effect of the azimuthal waves, such that the final resulting 
non-equilibrium appears completely manageable by the 14-moment model.

The 14-moment method appears to be, overall, at least 10 times more computationally demanding than the Euler equations,
but up to 1--100 times less expensive than a direct kinetic simulation.
These figures strongly depend on the number of cells, since fluid methods and the direct kinetic method scale with different powers
of the number of cells.
A comparison between the fluid and the kinetic methods is however not immediately obvious, especially when one employs a particle-based
approach, where the cost depends on the accepted level of statistical noise.
Such noise is known to have an important effect for self-consistent or collisional simulations.
Moreover, the computational cost of particle-grid mapping constitutes a significant part of the cost of a coupled simulation.
These points are however beyond the scope of this work.
Future studies should investigate the accuracy and efficiency of the 14-moment maximum-entropy method 
in fully coupled configurations, and investigate its ability to reproduce plasma oscillations and instabilities.


\section*{Acknowledgments}
We wish to thank A. Frezzotti (Politecnico di Milano) and F. Bariselli (Politecnico di Milano and von Karman Institute for Fluid Dynamics) for the enlightening discussions about this work.

This research was supported by grants from NVIDIA and utilized NVIDIA A100 40GB GPUs.
The authors are grateful for this support.

\appendix

\section{Full set of 14-moment equations}\label{appendix:full-eqs}

We provide here the full set of 14-moment equations. 
Index notation is employed here, with the convention that repeated indices imply summation.

\begin{subequations}
\begin{equation}
    \tfrac{\partial}{\partial t} \rho 
    + \tfrac{\partial}{\partial x_i} \left( \rho u_i \right) = S_1 
\end{equation}
\begin{equation}
    \tfrac{\partial}{\partial t} \left( \rho u_i \right)
    + \tfrac{\partial}{\partial x_j} \left( \rho u_i u_j + P_{ij} \right) = S_{2,3,4}
\end{equation}
\begin{multline}
    \tfrac{\partial}{\partial t} \left( \rho u_i u_j + P_{ij} \right)
    + \tfrac{\partial}{\partial x_k} \left( \rho u_i u_j u_k + u_i P_{jk} + u_j P_{ik} \right.\\
\left. + u_k P_{ij} + Q_{ijk} \right) = S_{5-10}
\end{multline}
\begin{multline}
    \tfrac{\partial}{\partial t} \left( \rho u_i u_j u_j + u_i P_{jj} + 2 u_j P_{ij} + Q_{ijj} \right) \\
    +\tfrac{\partial}{\partial x_k} \left(\rho u_i u_k u_j u_j  + u_i u_k P_{jj} + 2 u_i u_j P_{jk} + 2 u_j u_k P_{ij} \right. \\ 
    \left. + u_j u_j P_{ik} + u_i Q_{kjj} + u_k Q_{ijj} + 2 u_j Q_{ijk} + R_{ikjj} \right) = S_{11,12,13}
\end{multline}
\begin{multline}
    \tfrac{\partial}{\partial t} \left( \rho u_i u_i u_j u_j + 2 u_i u_i P_{jj} + 4 u_i u_j P_{ij} + 4 u_i Q_{ijj} + R_{iijj} \right) \\
    + \tfrac{\partial}{\partial x_k} \left( \rho u_k u_i u_i u_j u_j + 2 u_k u_i u_i P_{jj} + 4 u_i u_i u_j P_{jk} \right. \\
    \left. + 4 u_i u_j u_k P_{ij} + 2 u_i u_i Q_{jkk} + 4 u_i u_k Q_{ijj} + 4 u_i u_j Q_{ijk} \right. \\
    \left. + 4 u_i R_{ikjj} + u_k R_{iijj} + S_{kiijj} \right) = S_{14}
\end{multline}
\end{subequations}

\noindent For more details, see McDonald \& Torrilhon, \cite{mcdonald2013affordable} where an expression for the 
closing fluxes $Q_{ijj}$, $R_{ikjj}$ and $S_{kiijj}$ is also given.
For simplicity, these expressions are not reported here.


\section{Collisionless expansion of a quasi-neutral plasma}\label{app:examples-self-consistent}

As a complement to Section~\ref{sec:quasi-neutral-model}, we discuss here the classical problem of the planar 
expansion of a quasi-neutral collisionless plasma into a vacuum.\cite{mora2005collisionless}
Electrons are here assumed to be isothermal for simplicity, at $T_e = 20~\si{eV}$, constant and uniform throughout the simulation.
The initial plasma density is set to $n_{0} = 10^{17}~\si{m^{-3}}$, giving a Debye length of $\lambda_{\mathrm{De}} \approx 10^{-4}~\si{m}$.
Argon ions ($m_i=6.6337\times10^{-26}~\si{kg}$) are initialized with a zero velocity and with a Gaussian density profile:
\begin{equation}
  n(x, t=0) = n_{0} \exp(-x^2/R_0^2) \, ,
\end{equation}

\noindent with $R_0 = 0.05~\si{m}$. This satisfies the condition $R_0 \gg \lambda_{\mathrm{De}}$, 
and the plasma can be assumed to be quasi-neutral.
The simulated domain is centered in zero and has a total width of $L_x = 2~\si{m}$.
All other ion quantities are taken from a Maxwellian distribution at $T_i = 100~\si{K}$.
Such temperature is sufficiently low for ions to be considered cold, in the present problem.
With this choice, our results match the analytical solution.\cite{mora2005collisionless}

\begin{figure}[h!tb]
  \centering
  \includegraphics[width=1.0\columnwidth]{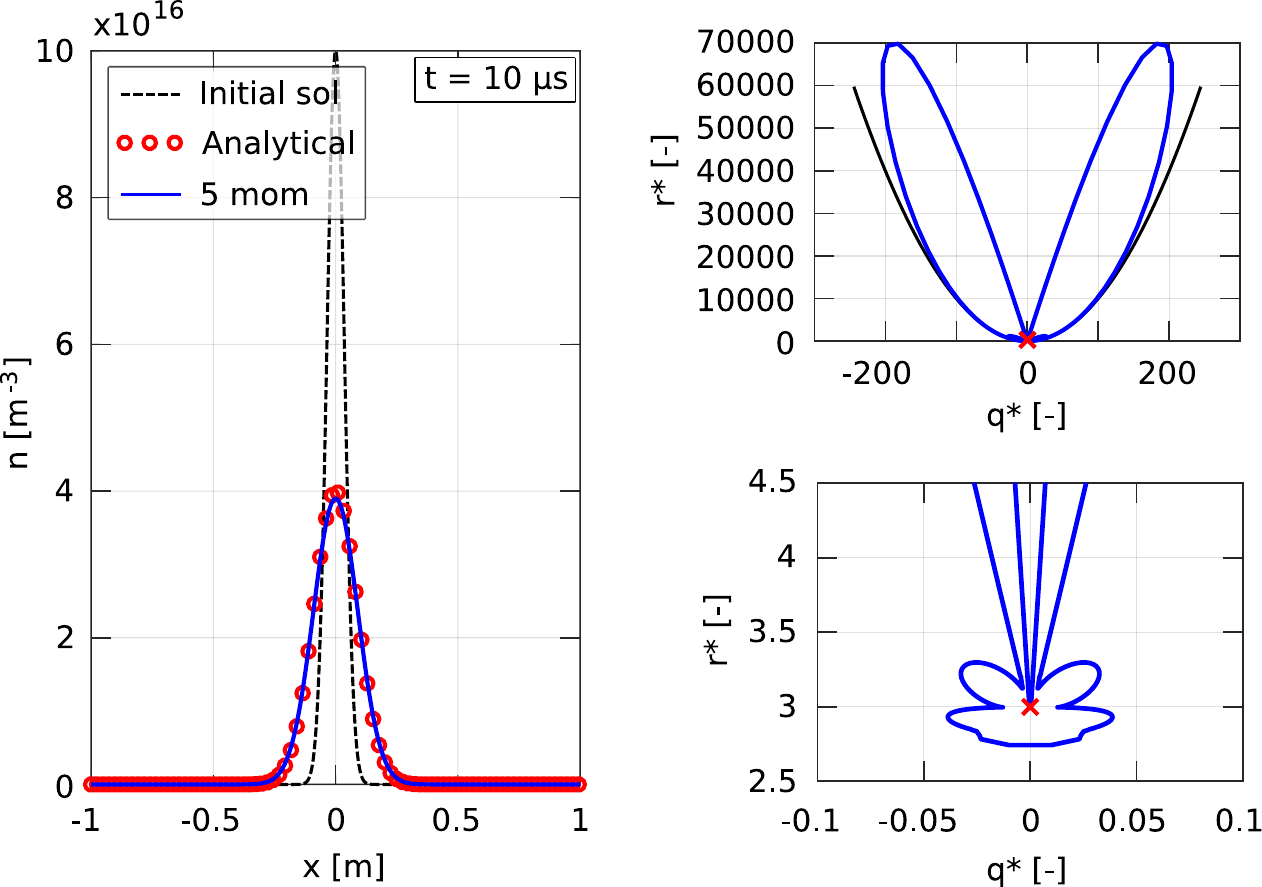}
  \caption{Solution of the 5-moment system for the planar quasi-neutral collisionless expansion test case. 
           Left: number density, compared to the analytical solution. Right: solution in dimensionless moment space.}
  \label{fig:quasi-neutral-test-case}
\end{figure}

The ion equations to be solved are Eqs.~\eqref{eq:5mom-sys}, 
with conserved variables $\bm{U}_5$ given by Eq.~\eqref{eq:U-5-definitions},
fluxes $\bm{F}_5$ from \eqref{eq:F-5-definitions}, no ionization source terms ($\bm{S}_{5}^{\mathrm{iz}}=0$), 
and quasi-neutral electrostatic sources $\bm{S}_5^{E}$ from \eqref{eq:quasi-neutral-sources-5mom}.
Figure~\ref{fig:quasi-neutral-test-case}-Left shows the resulting plasma density profile at a time $t = 10~\si{\mu s}$,
compared to the analytical results. 
As mentioned, the two models match since we are considering low-temperature ions.
Indeed, the pressure, as well as higher order central moments, scale with powers of the temperature, 
and thus have little effect on the density and the bulk velocity.
In other words, the thickness of the distribution function in velocity space is small with respect to the 
average velocity. 

For this specific problem, a moment method is not needed if one is only interested in the density and the bulk velocity.
Yet, for the sake of completeness, we report in Fig.~\ref{fig:quasi-neutral-test-case}-Right a plot of the solution 
in dimensionless moment space. 
The Right-Top panel shows the whole extent of the dimensionless solution, while the Right-Bottom panel shows 
a magnification around local thermodynamic equilibrium ($q^\star =0, r^\star =3$, identified by a red cross symbol).
Even for this simple problem, the VDF appears to assume extreme non-equilibrium states.

The computational cost for running this case appeared to be mostly connected to the source term, that is written in non-conservative form.
The wave speeds of the 5-moment system on the other hand remained affordably low during the simulation, as the 
solution does not cross the Junk subspace.
This may change in case warmer ions are considered.
However, this goes beyond the scope of this work.



\bibliographystyle{unsrt}

\end{document}